\documentclass[a4paper,fleqn,usenatbib]{mnras}

\usepackage{newtxtext,newtxmath}

\usepackage[T1]{fontenc}
\usepackage{ae,aecompl}
\usepackage{relsize} 
\usepackage{color,soul}
\usepackage{subfig}

\usepackage{graphicx}
\usepackage{amsmath}
\usepackage{bm}
\usepackage{footnote}
\usepackage{bigints}
\usepackage{etoolbox}
\usepackage{rotating}
\usepackage{tablefootnote}
\usepackage{threeparttable}
\usepackage{ulem}
\makeatletter
\AtBeginEnvironment{thebibliography}{%
   \clubpenalty10000
   \@clubpenalty \clubpenalty
   \widowpenalty10000}
\makeatother

\title[An upper limit for planet growth]{\vspace{-2mm}{An upper limit for the growth of inner planets?} \vspace{-3mm}}

\author[Winter \& Alexander]{Andrew J. Winter$^{1}$\thanks{andrew.winter@uni-heidelberg.de} and Richard Alexander$^{2}$
\\
$^{1}$Institut f\"{u}r Theoretische Astrophysik, Zentrum f\"{u}r Astronomie, Heidelberg University, Albert Ueberle Str. 2, 69120 Heidelberg, Germany \\
$^{2}$School of Physics and Astronomy, University of Leicester, Leicester, LE1 7RH, UK\\
}

\date{Accepted 2021 {May} 4. Received 2021 {May} 3; in original form March {2021} 26}\vspace{-2mm}

\pubyear{2021}

\begin{document}
\defcitealias{Mulders18}{M18}

\label{firstpage}
\pagerange{\pageref{firstpage}--\pageref{lastpage}}
\maketitle

\begin{abstract}
 The exotic range of known planetary systems has provoked an equally exotic range of physical explanations for their diverse architectures. However, constraining formation processes requires mapping the observed exoplanet population to that which initially formed in the protoplanetary disc. Numerous results suggest that (internal or external) dynamical perturbation alters the architectures of some exoplanetary systems. Isolating planets that have evolved without any perturbation can help constrain formation processes. We consider the \textit{Kepler} multiples, which have low mutual inclinations and are unlikely to have been dynamically perturbed. We apply a modelling approach similar to that of \citet{Mulders18}, additionally accounting for the two-dimensionality of the radius ($R_\mathrm{pl} =0.3{-}20\,R_\oplus$) and period ($P_\mathrm{orb} = 0.5-730$~days) distribution. We find that an upper limit in planet mass of the form $M_\mathrm{lim} \propto a_\mathrm{pl}^{\beta} \exp(-a_\mathrm{in}/a_\mathrm{pl})$, for semi-major axis $a_\mathrm{pl}$ and a broad range of $a_\mathrm{in}$ {and} $\beta$, can reproduce a distribution of $P_\mathrm{orb}$, $R_\mathrm{pl}$ that is indistinguishable from the observed distribution by our comparison metric. The index is consistent with $\beta= 1.5$, expected if growth is limited by accretion within the Hill radius. This model is favoured over models assuming a separable PDF in $P_\mathrm{orb}$, $R_\mathrm{pl}$. The limit, extrapolated to longer periods, is coincident with the orbits of RV-discovered planets ($a_\mathrm{pl}>0.2$~au, $M_\mathrm{pl}>1\,M_\mathrm{J}$) around recently identified low density host stars, hinting at isolation mass limited growth. We discuss the necessary circumstances for a coincidental age-related bias as the origin of this result, concluding that such a bias is possible but unlikely. We conclude that, in light of the evidence in the literature suggesting that some planetary systems have been dynamically perturbed, simple models for planet growth during the formation stage are worth revisiting.   
\end{abstract}

\begin{keywords} 
Planetary Systems -- stars: kinematics and dynamics -- circumstellar matter -- planets and satellites: formation\vspace{-2mm}
\end{keywords}


\section{Introduction}

Exoplanetary systems are diverse, with the constituent planets exhibiting a wide range of masses, semi-major axes and eccentricities \citep[see e.g.][for a review]{winn15}. Since the discovery of the first `hot Jupiter' orbiting 51 Pegasi at a separation of $\sim 0.05$ ~au \citep{Mayor95}, it has been clear that planet formation and evolution processes do not exclusively, or even typically, result in Solar system-like architectures. A wide variety of processes to explain this diversity have been suggested, pertaining both to the primordial protoplanetary disc of dust and gas from which the planets form \citep{Dullemond07, armitage11} and the subsequent evolution of the planetary system. These mechanisms are often strongly dependent on parameters that are poorly constrained empirically, such as disc viscosity \citep[e.g.][]{Bitsch13b} or the initial conditions for exoplanets undergoing dynamical evolution \citep[e.g.][]{Barnes06, Carrera19}. It is therefore challenging to demonstrate the relative importance of particular mechanisms in reproducing the observed exoplanet distribution.


A key step towards understanding the origin of the observed exoplanets and system architectures is separating out processes that govern isolated planet-disc evolution from those that may perturb the planetary system during or after formation. An illuminating example is the dichotomy between the single \textit{Kepler} planets and those found to be in multiple systems \citep{Lissauer11}. The single systems have higher eccentricities than those in multiple systems \citep{Xie16}, which may be the result of dynamical scattering and ejection or high mutual inclination of companions \citep{Zhu18}. Similarly, whether dynamical perturbation drives the inward migration of hot Jupiters remains a topic of debate \citep{dawson18}. {Their orbits may be the result of planet-planet \citep[e.g.][]{Weidenschilling96,Johansen12} or binary \citep[e.g.][]{Nagasawa08, Belokurov20-A} dynamical perturbation, or disc induced migration \citep[e.g.][]{Lin96, Heller19}.} 
Complicating things further, stars form in clusters or associations \citep{Lad03}, and planet formation is subject to numerous influences from its formation environment \citep{Parker20} that may leave an imprint on the planet population \citep{Winter20c,dai21,Adibekyan21,Longmore21}. Given these considerations, efforts to perform population synthesis for the observed population may be confounded by the fact that we are not observing a continuous population. This work is motivated by the growing evidence that a subset of planetary systems may have been perturbed from their `natural' formation configuration, and may therefore be confusing the interpretation of the exoplanet population as a whole.

In the remainder of this work, we explore the possibility that a simple model for the growth of planets may explain a subset of the observed exoplanet population that has not been perturbed. In Section~\ref{sec:theory_Miso} we outline the simplest theoretical expectation for a planet growing in a low viscosity disc. We then consider this limit in the context of two samples of planets that are the least likely to have undergone dynamical perturbation: \textit{Kepler} multiples (Section~\ref{sec:Kepmult}) and the recently identified RV planets hosted by stars with low position-velocity density (Section~\ref{sec:field_fit}). In this way, we demonstrate that an upper limit in planet growth may explain the properties of unperturbed systems. We discuss whether such a limit is consistent with other observational constraints in Section~\ref{sec:discuss}, and draw conclusions in Section~\ref{sec:concs}.

\section{Theoretical motivation}
\label{sec:isolation_mass}
\label{sec:theory_Miso}

A wealth of theoretical works have investigated the role of planet-disc interaction in inducing the migration of planets \citep{Kley12}. The main empirical motivation for such theory is the existence of massive planets at short orbital periods \citep[e.g.][]{Mayor95, dawson18}. However, the rate and stopping mechanisms \citep[e.g.][]{Rice08,  Ida08,Matsumura09}, or even the direction \citep[e.g.][]{Michael11}, of migration are all sensitive to underlying assumptions. The outcome of migration is thus dependent on many parameters that have relatively poor empirical constraints, and affect both the initial conditions and disc physics. Mounting evidence suggests that some combination of environment \citep{Brucalassi16,shara16-A, Winter20c, dai21}, multiplicity \citep{Belokurov20-A, Hirsch21}, or orbital instability \citep{Weidenschilling96, Johansen12}, leaves an imprint on planetary systems.  It is therefore possible that the torques exerted on a planet during formation do not lead to efficient migration (i.e. from several tens to $\sim 0.1$~au).  In this case, complex formation models applied to explain the population of observed exoplanets may not be necessary. To explore this possibility, it is instructive to consider the limit of no migration; a simple model in which the planet simply grows by accreting in its vicinity.

The `isolation mass' is the mass to which a planet can grow by clearing out an annulus of material within the protoplanetary disc. This mass is determined by the local surface density $\Sigma_\mathrm{pl}$ at the location of the planet (or planetesimal). While the isolation mass is usually considered in the context of initial accretion of solid material (dust), the same principles can apply to gas accretion if angular momentum transport in the disc is inefficient. This is the simplest picture of planet growth, where gas transport across the initial gap opened up by the nascent planet is negligible (i.e. the case of low viscosity). {If the growth of a planet is not limited by the opacity of accreting material, then runaway growth is possible} \citep[e.g.][]{Pollack96}. In this case, the isolation mass can be estimated by assuming that planetesimals initially follow circular orbits, and accrete all material approximately within their Hill radius $r_\mathrm{H}$. The resultant isolation mass estimate is:
\begin{align}
\label{eq:Miso1}
\nonumber
    M_\mathrm{iso}& \approx 2 \pi \int_{a_\mathrm{pl}-\Delta a}^{a_\mathrm{pl}+\Delta a } \! \Sigma(r)\, r \, \mathrm{d} r.
\end{align} 
where $a_\mathrm{pl}$ is the the semi-major axis of the forming planet. The annulus radius is approximated $\Delta a \approx  C r_\mathrm{H}$, where the constant $C\approx 2\sqrt{3}$ \citep{Lissauer93, armitage2007lecture}. One then needs to make assumptions about the functional form of the surface density, which we will assume has the form (see discussion in Section~\ref{sec:discuss_massviscfeed}):
\begin{equation}
\label{eq:Sigma}
    \Sigma(a) = \Sigma_0  \frac{m_*}{m_{*,0
    }} \left( \frac{a}{a_0}\right)^{-\gamma}, 
\end{equation} {where subscript `$0$' values refer to normalisation constants in units of the associated variables.} In this case the isolation mass is determined by the equation:
\begin{align}
\nonumber
    M_\mathrm{iso} \approx &\frac{2\pi \Sigma_0}{2-\gamma} \frac{m_*}{m_{*,0
    }} a_0^\gamma a_\mathrm{pl}^{2-\gamma}\times \\& \times \left\{\left[1+C\left(\frac{M_\mathrm{iso}}{m_*}\right)^{1/3}\right]^{2-\gamma} -\left[1-C\left(\frac{M_\mathrm{iso}}{m_*}\right)^{1/3}\right]^{2-\gamma}\right\},
    \label{eq:Miso_def}
\end{align} with no analytic solution for general $\gamma$. 

{For a viscous disc, the radial dependence of the disc surface density is dependent on the temperature profile. Since we are assuming little migration, we expect to be in the regime of low viscosity, where heating is radiative rather than viscous. For a constant viscous $\alpha$-disc that is not viscously heated, the  mid-plane temperature is $T\propto a^{-1/2}$ {\citep{Kenyon87}}, for radius $a$ within the disc. In equilibrium, the corresponding surface density in the intermediate region between the inner and outer edges of the disc is $\Sigma \propto a^{-1}$ \citep[$\gamma=1$ --][]{Har98}.}  Hence equation~\ref{eq:Miso_def} becomes:
\begin{align}
    \label{eq:Miso_units}
\nonumber
   M_\mathrm{iso} &\approx \frac{8}{\sqrt{3}} \pi^{3/2} C^{3/2} m_*^{-1/2} \Sigma_\mathrm{pl}^{3/2} a_\mathrm{pl}^3. \\
   & = 55.9  \left( \frac{\Sigma_0}{10^3 \, \mathrm{g ~
    cm}^{-2}}\right)^{3/2}   \left( \frac{m_*}{1\, M_\odot}\right) \left( \frac{a_\mathrm{pl}}{1\, \mathrm{au}}\right)^{3/2} \, M_\oplus,
\end{align} where we have chosen $m_{*,0} = 1\, M_\odot$, $a_0 = 1$~au. 

\begin{figure}
    \centering
    \includegraphics[width=0.95\columnwidth]{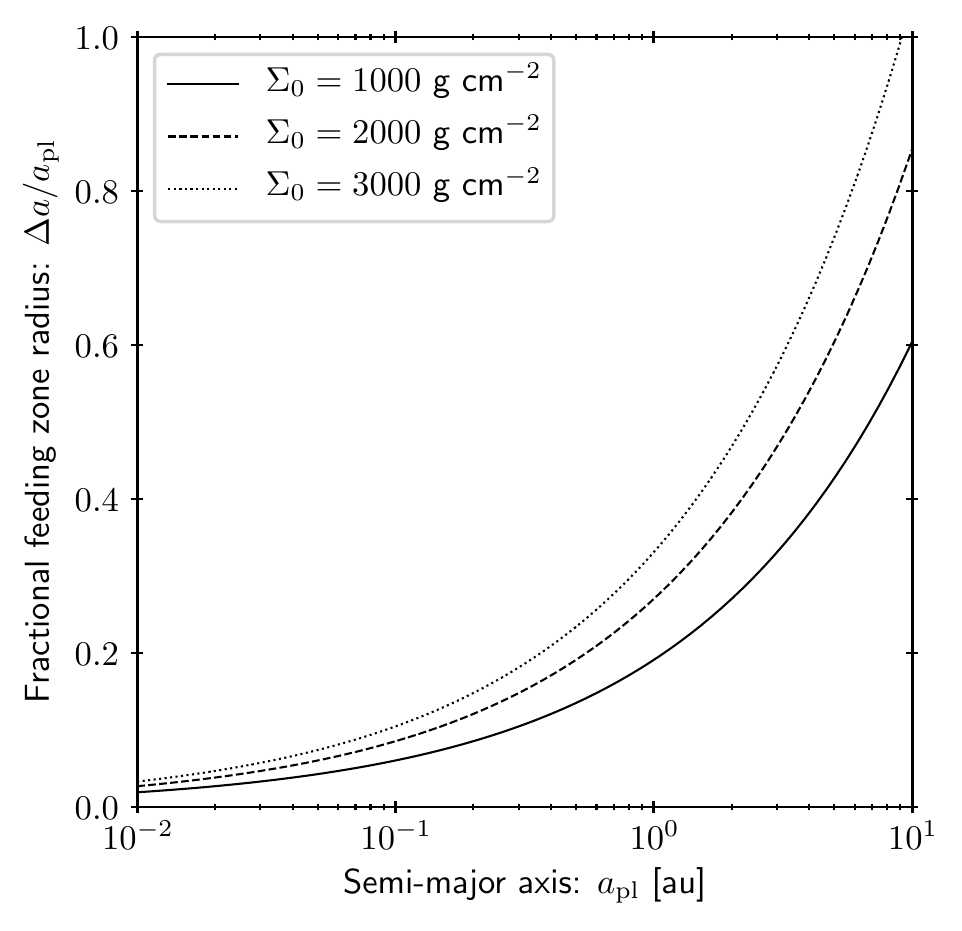}
    \caption{Feeding zone radius $\Delta a = C r_\mathrm{H}$ for $r_\mathrm{H}$ the Hill radius as a fraction of planet semi-major axis $a_\mathrm{pl}$ based on equation~\ref{eq:Miso_units}.  }
    \label{fig:fracfeed}
\end{figure}

However, equation~\ref{eq:Miso_units} cannot apply throughout the disc, since as the semi-major axis $a_\mathrm{pl}$ becomes large, the fractional feeding zone radius $\Delta a/a_{\rm{pl}}$ also increases. When $\Delta a \rightarrow a_\mathrm{pl}$, the approximation must break down, and the feeding zone radius is no longer proportional to the Hill radius. We show $\Delta a/a_{\rm{pl}}$ as a function of $a_{\rm{pl}}$ in Figure~\ref{fig:fracfeed}, illustrating that we expect the small feeding zone approximation to break down in the range $a_{\rm{pl}}\sim 1{-}10$~au. Because we have assumed that surface density scales linearly with stellar host mass $m_*$, $\Delta a/a_{\rm{pl}}$ is independent of $m_*$. If planet growth is limited by the isolation mass we would expect an upper limit following equation~\ref{eq:Miso_units} for planets with semi-major axes $a_\mathrm{pl}\lesssim 1$~au, transitioning to a shallower relationship at some larger $a_\mathrm{pl}$.

\section{Upper mass limit for \textit{Kepler} multiples}
\label{sec:Kepmult}
Due to its well-characterised detection efficiency, data from the \textit{Kepler} mission has provided invaluable constraints on the occurrence rates of exoplanets within $\sim 1$~au of their host star \citep[e.g.][]{Borucki16}. However, inferring occurrence rates from the full sample suffers from a potential limitation similar to that of population synthesis models; the implicit assumption that a single continuous sample of exoplanets are observed. In particular, the observed dichotomy of single and multiple systems in the \textit{Kepler} sample \citep{Lissauer11} is strong evidence that the observed exoplanets form through multiple distinct channels. The higher eccentricities of single planets \citep{Xie16, Mills19, Bachmoller20} suggests that an internal \citep{Carrera19} or external \citep{dai21,Rodet21,Longmore21} perturbation has driven scattering and high eccentricities, either ejecting planets or yielding high mutual inclinations \citep{Zhu18}. By contrast, multiple systems discovered by transit must have similar inclinations and are therefore most likely to have formed without a dynamical perturbation. To quantify the relative occurrence rates of planetary systems forming in isolation, it is therefore instructive to consider only the \textit{Kepler} multiple systems. 

Here we consider the probability density function (PDF) of exoplanet properties as a function of mass and semi-major axis. Our approach is inspired by that of \citet[][hereafter \citetalias{Mulders18}]{Mulders18}, with a number of important differences. First, we are not interested in absolute occurrence rates, only the relative distribution of detected exoplanets. We therefore need only consider confirmed exoplanet detections, as opposed to all \textit{Kepler} targets. Related to this point, we restrict ourselves to \textit{Kepler} multiple systems for the reasons outlined above. In addition, we do not assume that the two dimensional PDF is a separable function of period and radius. Specifically, we will ask whether an upper limit in exoplanet masses is statistically favoured over the separable broken power-law model adopted by \citetalias{Mulders18}. Following from this, the metric we adopt for comparison between observations and models accounts for the covariance of the period-radius distribution, rather than assuming that period and radius distributions can be compared independently. Finally, our PDF is defined in terms of mass and semi-major axis rather than radius and orbital period. This is significant for the functional form of our PDF, but we convert to period and radius using empirical constraints for comparison with the observed distribution (Section~\ref{sec:MR_relation}). 

In the remainder of this section we describe the sample and cuts made in Section~\ref{sec:Kepsample}, then the model for the PDF in Section~\ref{sec:model}. The (relative) completeness of the \textit{Kepler} sample is described in Section~\ref{sec:Kep_completness}, and the models we evaluate are outlined in Section~\ref{sec:model}. The parameter space exploration is described in Section~\ref{sec:param_exp}. Our results are summarised in Section~\ref{sec:Kep_results}.

\subsection{Sample}
\label{sec:Kepsample}

\begin{figure*}
    \centering
    \includegraphics[width=\textwidth]{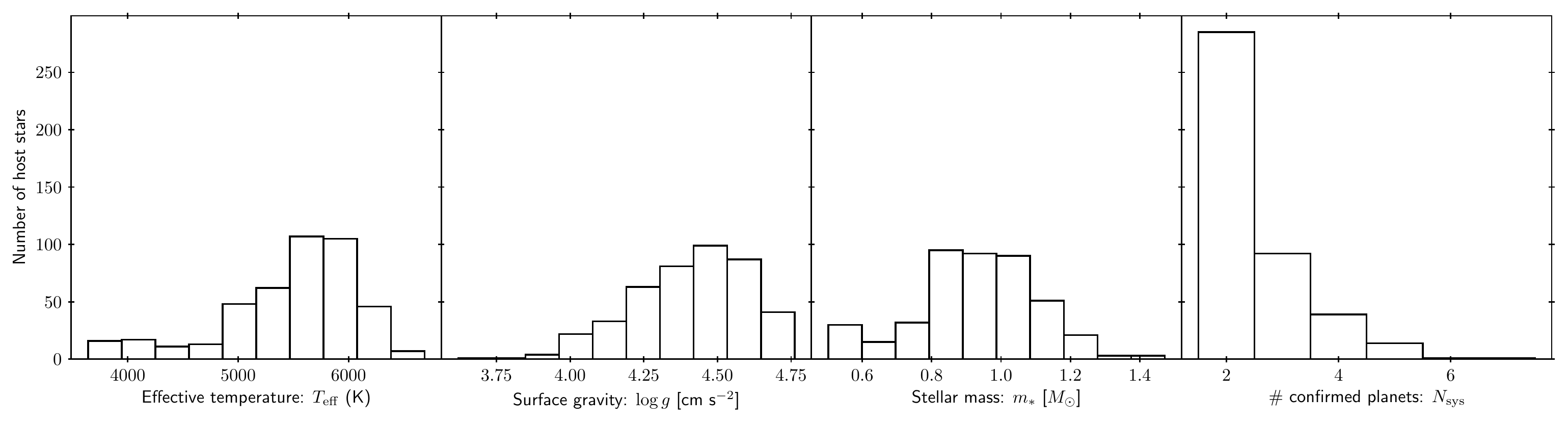}
    \caption{Histograms for the properties of the host stars of the fiducial sample of exoplanetary systems used in Section~\ref{sec:Kepmult}. The properties left-most properties from left to right are the stellar effective temperature, surface gravity and mass. The histogram on the right is the number of detected transiting planets in the system. Sample includes only systems with two or more confirmed transiting planets.  }
    \label{fig:Kephosts}
\end{figure*}

Throughout this section we use the exoplanet and stellar host properties from the \textit{Kepler} Data Release 25 (DR25), which is a uniform processing of the full four year dataset \citep[Q1-17 -- e.g.][]{Twicken16, Berger18}. These data were downloaded from the \citeauthor{exoarchive} \citep[NEA --][]{Akeson13} on the 29$^{\mathrm{th}}$ November 2020. As discussed above, we are interested in recovering the distribution of exoplanets in systems that are the least likely to have undergone a perturbation subsequent to formation. We therefore consider only systems with two or more confirmed transiting exoplanet detections, implying low mutual inclinations. This choice has the added benefit that false positives are unlikely. Since we expect an upper mass limit to be dependent on stellar mass, we wish to consider a narrow range of estimated host star masses $0.5{-}1.5 \, M_\odot$. We further exclude evolved stars with surface gravity $\log g <3.5$ (where $g$ is in cgs units). The properties of the final host star sample are shown in Figure~\ref{fig:Kephosts}. It comprises 432 systems, of which 285 have two detected planets and 92 have three. The total number of planets in these systems is 1085.

\subsection{Survey completeness}
\label{sec:Kep_completness}

\begin{figure}
    \centering
    \includegraphics[width=0.95\columnwidth]{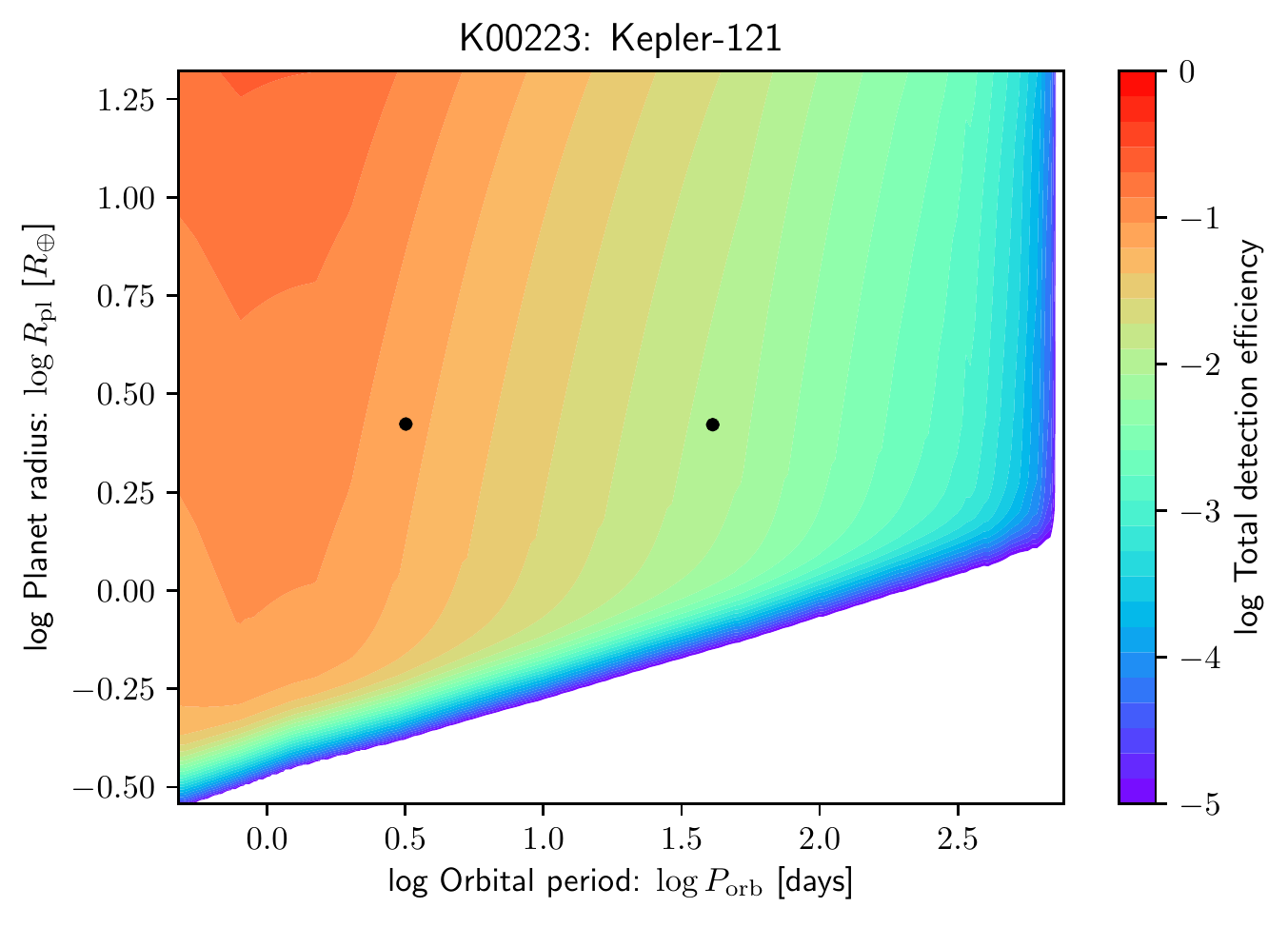}
    \caption{Total detection efficiency for the host star Kepler-121. The colour bar shows the fraction of planets at a given orbital period ($0.5{-}730$~days) and planet radius ($0.3{-}20\, R_\oplus$) and random inclination that would be considered a detection as a consequence of geometric effects and processing by the pipeline summarised in Section~\ref{sec:Kep_completness}. Black points mark the measured periods and radii of Kepler-121 b and c. }
    \label{fig:detcont_eg}
\end{figure}

In order to draw comparisons between relative exoplanet frequencies in our simple model and those of the \textit{Kepler} planets, we need an estimate of the relative fraction of planets at a given mass and semi-major axis that would be detected by the recovery pipeline -- i.e. the relative false negative frequency. For a fuller discussion of the completeness considerations, see \citetalias{Mulders18}; we briefly summarise as follows. 

The most obvious bias in the transit detection efficiency is the geometric probability that a planet will transit. For a circular orbit, this probability is:
\begin{equation}
    f_\mathrm{geo} = 0.0046 \left(\frac{a_\mathrm{pl}}{1\, \mathrm{au}} \right)^{-1} \frac{R_*+R_\mathrm{pl}}{R_\odot},
\end{equation} corresponding to reduced detection efficiencies at long orbital periods.

At short orbital periods $\lesssim 3$~days further incompleteness factor $f_\mathrm{harm}$ is due to the behaviour of the harmonic fitter, applied to remove periodic stellar activity in the {Transiting Planet Search} \citep[TPS --][]{Jenkins10, Tenenbaum12,Christiansen13}. We interpolate the false negative rates between $0.5{-}3$~days estimated by \citet{Christiansen15}, adopting values of $f_\mathrm{harm}$: $0.4$ at $0.5$~days, $0.6$ at $0.8$~days, $0.9$ at $1.5$~days, and $0.99$ at $2.0$~days, linearly interpolating between these values in this range. 

The completeness for a given data release from the \textit{Kepler} satellite is limited at long periods by the requirement of three transits across the baselines -- i.e. a maximum orbital period of two years for the entire four year dataset. In addition, for small planet radii $R_\mathrm{pl}$, the chance of detection is limited by the signal-to-noise ratio. Factoring in these variable detection efficiencies, the completeness of the \textit{Kepler} transiting sample can be quantified as a function of the {Multiple Event Statistic} (MES) used in the TPS algorithm. We calculate the signal-to-noise detection efficiency ($f_{\mathrm{S/N}}$) for each star $i$ in our sample using \textsc{KeplerPORTs}\footnote{\url{https://github.com/nasa/KeplerPORTs}} \citep{Burke17}.

Finally, even detectable exoplanets are not necessarily vetted as planet candidates by the \textit{Kepler Robovetter} \citep{Coughlin16}. We use the parametric approximation for the vetting efficiency by \citetalias{Mulders18}, based on the simulated data products\footnote{\url{https://exoplanetarchive.ipac.caltech.edu/docs/KeplerSimulated.html}} and following \citet{Thompson18}:
\begin{equation}
    f_\mathrm{vet} =
    c \left(\frac{R_\mathrm{pl}}{R_\oplus}\right)^{\alpha_R} 
\begin{cases}
 \left(P_\mathrm{orb}/P_\mathrm{break}\right)^{\alpha_P}  \qquad P_\mathrm{orb}<P_\mathrm{break}\\
  \left(P_\mathrm{orb}/P_\mathrm{break}\right)^{\beta_P} \qquad P_\mathrm{orb} \geq P_\mathrm{break}
\end{cases}
\end{equation} with $c = 0.63$, $\alpha_R = 0.19$, $P_\mathrm{break} = 53$~days, $\alpha_P = -0.07$,
$\beta_P = -0.39$.

The overall detection efficiency for a given host star $i$ is then given by:
\begin{equation}
    f_\mathrm{det}(P_\mathrm{orb},R_\mathrm{pl}; i) = f_\mathrm{geo}\cdot f_\mathrm{harm} \cdot f_{\mathrm{S/N}} \cdot f_\mathrm{vet}.
\end{equation} We show an example of the total detection efficiency for the host star Kepler-121 in Figure~\ref{fig:detcont_eg}. 

\subsection{Model probability density functions}
\label{sec:model}
\subsubsection{Broken power-law model}

\citetalias{Mulders18} made the simplifying assumption that the logarithmic PDF that describes the relative number $N_\mathrm{pl}$ of the \textit{Kepler} planets at a given period and radius {is} a separable function of period and radius:
\begin{equation}
    f_\mathrm{M18} \equiv \frac{\mathrm{d}^2 N_\mathrm{pl}}{\mathrm{d} \log P_\mathrm{orb}\, \mathrm{d} \log R_\mathrm{pl}}  \propto f_R(R_\mathrm{pl}) \cdot f_P(P_\mathrm{orb}).
\end{equation}This PDF is written as the product of two broken power-laws:
\begin{equation}
    f_P = 
\begin{cases}
 \left(P_\mathrm{orb}/P_\mathrm{break}\right)^{a_P} & \qquad P_\mathrm{orb}<P_\mathrm{break}\\
  \left(P_\mathrm{orb}/P_\mathrm{break}\right)^{b_P}& \qquad P_\mathrm{orb} \geq P_\mathrm{break}
\end{cases}
\end{equation} and
\begin{equation}
    f_R = 
\begin{cases}
 \left(R_\mathrm{pl}/R_\mathrm{break}\right)^{a_R} & \qquad R_\mathrm{pl}<R_\mathrm{break}\\
  \left(R_\mathrm{pl}/R_\mathrm{break}\right)^{b_R} &\qquad R_\mathrm{pl} \geq R_\mathrm{break}
\end{cases},
\end{equation} for which the constants $a_P$, $b_P$, $P_\mathrm{break}$, $a_R$, $b_R$ and $R_\mathrm{break}$ are free parameters. \citetalias{Mulders18} also estimated the absolute occurrence rate (normalisation), or number of planets per star. However, we are here only interested in the relative planet numbers, since the motivation for this work is that some subset of planets are a distinct population in the first instance. This leaves six parameters for which to fit. 

\subsubsection{Mass limited model}

We consider an alternative to the above model:
\begin{equation}
    f_\mathrm{iso}(a_\mathrm{pl}, M_\mathrm{pl}) \equiv \frac{\mathrm{d}^2 N_\mathrm{pl}}{\mathrm{d} \log a_\mathrm{pl} \, \mathrm{d}\log M_\mathrm{pl}}\propto  f_\mathrm{lim}(M_\mathrm{pl}; a_\mathrm{pl}) \cdot f_M (M_\mathrm{pl}).
\end{equation} In this case, the PDF {of} planet masses follows a similar broken power-law as before:
\begin{equation}
    f_M = 
\begin{cases}
 \left(M_\mathrm{pl}/M_\mathrm{break}\right)^{a_M} & \qquad M_\mathrm{pl}<M_\mathrm{break}\\
  \left(M_\mathrm{pl}/M_\mathrm{break}\right)^{b_M}& \qquad M_\mathrm{pl} \geq M_\mathrm{break}
\end{cases}, 
\end{equation} which is empirically motivated \citep{Howard12, Petigura13}, and possibly reflects two physical regimes in the growth of planets. However, the overall two-dimensional PDF is no longer a separable function of semi-major axis (period) and mass (radius). The PDF $f_\mathrm{lim}(M_\mathrm{pl}; a_\mathrm{pl})$ instead represents a `cusp', or upper limit in planet mass $M_\mathrm{pl}$ that is dependent on their semi-major axis $a_\mathrm{pl}$.

Unlike the (double) broken power-law model, the model presented here does not explicitly treat the observed break in planet frequency at $P_\mathrm{orb}\sim 10$~days \citep{Howard12}. However, the PDF is not separable into components that are functions of only $M_\mathrm{pl}$ (or $R_\mathrm{pl}$) and $a_\mathrm{pl}$ (or $P_\mathrm{orb}$). Thus a break in the PDF of $M_\mathrm{pl}$ also produces a break in the PDF of $a_\mathrm{pl}$. 

If disc viscosity is low and type II planet migration is inefficient, then in situ growth after gap opening eventually leads to the exhaustion of material that the planet(esimal) can easily accrete. As discussed in Section~\ref{sec:theory_Miso}, such a mass can be approximated by estimating the quantity of disc material that orbits within the planet's Hill radius. Planet mass should therefore be limited by the isolation mass:
\begin{equation}
    M_\mathrm{lim} = M_\mathrm{iso}.
\end{equation}In deriving the isolation mass in Section~\ref{sec:theory_Miso}, we neglected the deviations in the disc surface density at the inner edge. Given that the \textit{Kepler} planets are close-in, we might expect the surface density to steepen in this regime. We therefore write the mass limit:
\begin{equation}
\label{eq:Mlim}
    M_\mathrm{lim} = M_0\cdot \exp\left( -\frac{a_\mathrm{in}}{a_\mathrm{pl}}\right) \cdot \left(\frac{a_\mathrm{pl}}{1\, \mathrm{au}}\right)^{\beta} ,
\end{equation} where $a_{\mathrm{in}}\propto R_\mathrm{in}$, the inner edge of the disc. Strictly, $\beta$ is not a free parameter since $\beta=1.5$ is a consequence of the assumptions discussed in Section~\ref{sec:theory_Miso}. We therefore initially consider only the normalisation constant $M_0$ and $a_\mathrm{in}$ as free parameters, although we subsequently also fit for $\beta$.

Defining the mass limit as in equation~\ref{eq:Mlim}, we can impose a smooth decline in the probability of a planet having mass $M_\mathrm{pl}$ above this limit:
\begin{equation}
    f_\mathrm{lim}(M_\mathrm{pl}; a_\mathrm{pl}) = \begin{cases}
     (2 \pi \sigma_M^2)^{-1/2}  &\qquad M_\mathrm{pl}<M_\mathrm{lim}\\
      f_\mathrm{ln}(M_\mathrm{pl}/M_\mathrm{lim};\sigma_M)& \qquad M_\mathrm{pl}\geq M_\mathrm{lim}
    \end{cases},
\end{equation} where 
\begin{equation}
     f_\mathrm{ln} (X; \sigma_X) = \frac{1}{\sigma_X\sqrt{2\pi }} \exp \left( -\frac{[\ln X]^2}{2\sigma_X^2}\right), 
\end{equation} is the log-normal PDF in log-space. The dispersion $\sigma_M$ also implicitly accounts for any dispersion between the mass and radius conversion. We have therefore introduced three additional free-parameters $M_0$, $a_\mathrm{in}$ and $\sigma_M$ that describe the hypothesised upper limit in planet masses, for a total of six free parameters with $a_M$, $b_M$ and $M_\mathrm{break}$.

\subsubsection{Conversion to observable quantities}
\label{sec:MR_relation}
The radius and period of a given planet are the quantities that can be inferred directly from {transit} measurements, while our model is a prediction for the mass and semi-major axis distributions. We therefore need to convert the physical quantities predicted by the model to the observable ones. To convert semi-major axes to orbital period, we simply adopt a stellar mass $m_*=1\, M_\odot$. Converting mass to radius, however, requires appealing to empirical constraints.

We convert exoplanet masses into radii using the recently inferred mass-radius relationship \citep{Otegi20}:
\begin{equation}
\label{eq:MR_relation}
    \frac{R_\mathrm{pl}}{R_\oplus} =\begin{cases}
 \mathcal{R}_1 (M_\mathrm{pl}) & \qquad M_\mathrm{pl}<3.11 \, M_\oplus \\
 \mathcal{R}_2(M_\mathrm{pl}) & \qquad 3.11 \, M_\oplus\leq M_\mathrm{pl}<92.2\, M_\oplus \\
\mathcal{R}_3(M_\mathrm{pl})  &\qquad M_\mathrm{pl}\geq 92.2\, M_\oplus
\end{cases},
\end{equation} {where
\begin{align}
\nonumber
    \mathcal{R}_1 &\equiv 1.03 (M_\mathrm{pl}/M_\oplus)^{0.29} \\
    \label{eq:RM_parts}
    \mathcal{R}_2 &\equiv 0.70
    (M_\mathrm{pl}/M_\oplus)^{0.63} \\
    \nonumber
    \mathcal{R}_3 &\equiv 11.6  (M_\mathrm{pl}/M_\oplus)^{0.01}. 
\end{align}}The threshold of the high mass regime is chosen such that the relation inferred for the intermediate mass regime, intersects the shallow high mass relationship at $R_\mathrm{pl}=12.1\, R_\oplus$ \citep{Bashi17}. The PDF in $\log R_\mathrm{pl}$, $\log P_\mathrm{orb}$ space is then:
\begin{equation}
\label{eq:pdf_conversion}
    f'_\mathrm{iso}\propto \frac{\mathrm{d}\log M_\mathrm{pl}}{\mathrm{d}\log R_\mathrm{pl}} \frac{\mathrm{d}\log a_\mathrm{pl}}{\mathrm{d}\log P_\mathrm{orb}} f_\mathrm{iso},
\end{equation} normalised over the domain of interest. 

{Strictly speaking, the physical relationship is not a well-defined mapping from mass to radius. This is because the rocky (small radius) and the gaseous (larger radius) populations overlap in mass in the range $M_\mathrm{pl}\sim 3 {-} 30\, M_\oplus$ \citep{Otegi20}. At planet masses in this range, a planet may retain a (tenuous) gaseous envelope or this envelope may be photoevaporated \citep{Owen12, Owen13}.  The transition is responsible for the `Fulton gap', an absence of planets with radii $R_\mathrm{pl}\sim 2 \, R_\oplus$ \citep{Fulton17}. In principle, one could account for the overlap by estimating the ratio of rocky to gaseous planets. However, this would complicate the model, especially in the absence of any strong constraints for the ratio. We therefore adopt equation~\ref{eq:MR_relation} in converting masses and radii in our fiducial model.} 
\begin{figure}
    \centering
    \includegraphics[width=0.95\columnwidth]{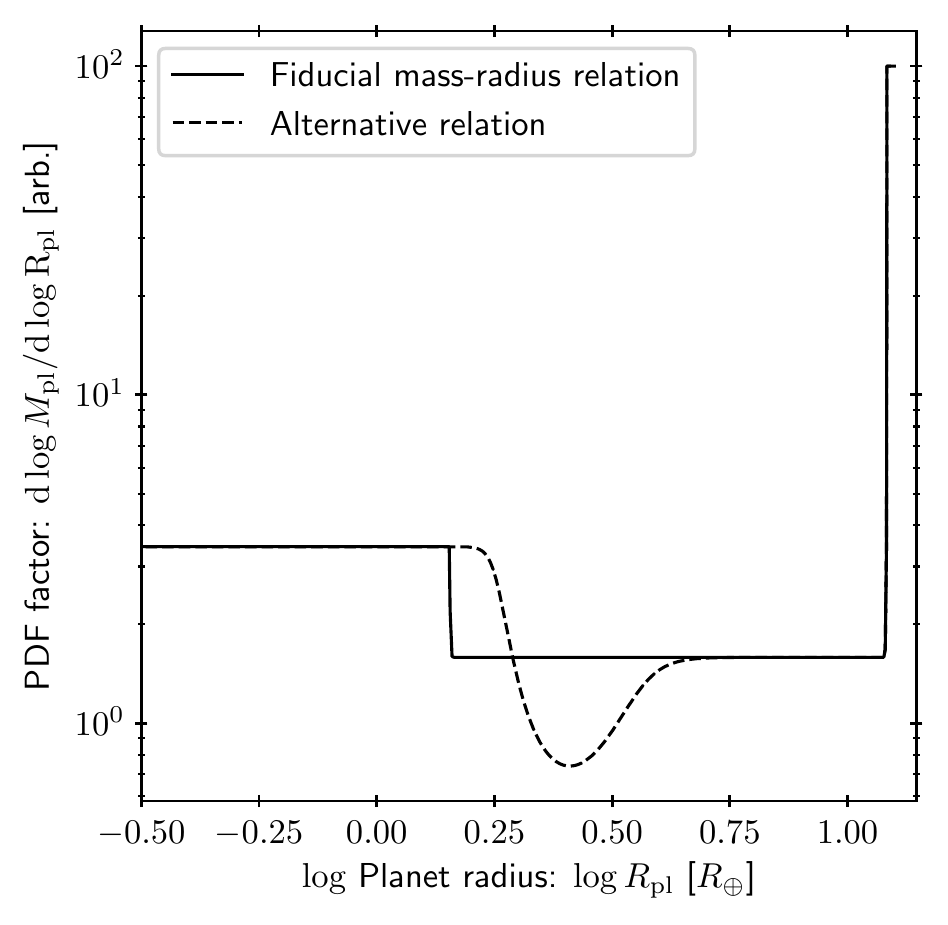}
    \caption{{The factor applied to the two-dimensional PDF in period-radius space (equation~\ref{eq:pdf_conversion}) using the two different planet mass-radius relations we adopt, as a function of planet radius and in arbitrary units. For the fiducial relationship (equation~\ref{eq:MR_relation}, solid line), this factor is proportional to the inverse of the power-law index for the local relation $\mathcal{R}_{1,2,3}(M_\mathrm{pl})$ in radius space. By contrast, the alternative relation (equation~\ref{eq:MR_Fulton}, dashed line) incorporates a smooth transition between rocky and gaseous relations at higher planet mass (radius).} }
    \label{fig:dMdR_conv}
\end{figure}

{We investigate the influence of our choice of mass-radius relation by considering an alternative relation in which planets are predominantly rocky ($R_\mathrm{pl} \propto \mathcal{R}_{1}$) up to a greater mass than assumed in equation~\ref{eq:MR_relation} ($M_\mathrm{pl}>3.11\, M_\oplus$). However, such an exercise requires revisiting the functional form of the mass-radius relation. While strictly equation~\ref{eq:MR_relation} is not differentiable at the hard transitions between mass-radius relations, the coordinate transform in equation~\ref{eq:pdf_conversion} is numerically calculable over our grid because the transitions occur at infinitesimal ranges in planet radius space. However, here we cannot allow a hard transition ($R_\mathrm{pl}\propto \mathcal{R}_1 \rightarrow R_\mathrm{pl}\propto\mathcal{R}_2$) because there exists a (non-negligible) discontinuity in the relation as a function of radius. This would result in an undefined PDF over the gap when applying the co-ordinate transformation (equation~\ref{eq:pdf_conversion}). }

{To account for the above numerical concerns, and in order to explore the influence of an alternative mass-radius relation, we adopt:
\begin{equation}
\label{eq:MR_Fulton}
    \left.\frac{R_\mathrm{pl}^\mathrm{rg}}{R_\oplus} \right. =\begin{cases}
\left[1-\mathcal{W}_\mathrm{rg}
\right]\mathcal{R}_1  + \mathcal{W}_\mathrm{rg} \cdot \mathcal{R}_2 &\quad M_\mathrm{pl}<92.2\, M_\oplus\\
\mathcal{R}_3  &\quad M_\mathrm{pl}\geq 92.2\, M_\oplus
\end{cases},
\end{equation} where 
\begin{equation}
    \mathcal{W}_\mathrm{rg}(M_\mathrm{pl}) = \frac 1 2 \left[ 1+ \tanh \left(\frac{\log {M_\mathrm{pl}/M_\mathrm{rg}}}{\Delta_\mathrm{rg}} \right) \right], 
\end{equation} such that $R_\mathrm{pl}^\mathrm{rg}(M_\mathrm{pl})$ is continuous and differentiable over the rocky-gaseous transition. Here $M_\mathrm{rg}$ and $\Delta_\mathrm{rg}$ are parameters that set the location and width of the transition from rocky to gaseous mass-radius relations respectively. In principle, we could allow $M_\mathrm{rg}$ and $\Delta_\mathrm{rg}$ to vary as free parameters. However, we will find that the best fitting fiducial model with six free parameters is indistinguishable from the observed data by our comparison metric. It is therefore not appropriate to attempt fitting two additional parameters. Instead we fix $M_\mathrm{rg} = 10\, M_\oplus$ and $\Delta_\mathrm{rg}=  0.1$, resulting in a transition between rocky and gaseous planets at approximately the mid-way mass in the overlap in the $\mathcal{R}_1$ and $\mathcal{R}_2$ regimes. While this choice is semi-arbitrary, it serves the purpose of investigating the influence of our choice of mass-radius relation on our fitting results. In Figure~\ref{fig:dMdR_conv} we show the factors applied to the two-dimensional PDF due to the adopted mass-radius relations. }

\subsection{Parameter space exploration}
\label{sec:param_exp}
\subsubsection{Sampling}

To assess the likelihood of a given model described by parameters $\Theta$, we need to compare to the observed sample $\zeta$. To achieve this, we first define the two dimensional model PDF across a grid in $P_\mathrm{orb}{-}R_\mathrm{pl}$ space. We choose a grid log-uniformly spaced between  $0.5{-}730$~days and $0.3{-}20\,R_\oplus$ at resolution $200$ in period and $150$ in radius. We draw a sample of planets $\tilde{\zeta}_\mathrm{all}$ from this grid. To ensure synthetic quantities are not fixed to a grid cell value, we offset each period and radius by an amount drawn from a uniform distribution between plus/minus half a grid cell. We thus obtain a period $P_i$ and radius $R_i$ for each synthetic planet.

To directly compare the model results with the observations, we must now account for the \textit{Kepler} selection biases. We select a subset of these planets by Monte Carlo sampling from $\tilde{\zeta}_\mathrm{all}$ with probabilities proportional to detection efficiencies at $P_i$, $R_i$. 
The specific detection efficiencies $\mathcal{D}_i$ for each synthetic planet $i$ are drawn at random from the pre-computed grid of the observed sample of host stars. This grid has the same resolution as the model PDF. We then draw a random number $u_i \sim \mathcal{U}(0,1)$ for each synthetic planet, and consider it to be `detected' for $u_i< \mathcal{D}_i /\rm{max} \mathcal{D}_i$. This yields a new sample $\tilde{\zeta}_\mathrm{det}$. Unlike \citetalias{Mulders18}, we are aiming only to quantify relative occurrence rates, and therefore do not need to reproduce the absolute number of `detected' planets. We are therefore free to normalise the $\mathcal{D}_i$ by the maximum detection efficiency $\rm{max} \mathcal{D}_i$, in order to maximise the size of the detected sample $|\tilde{\zeta}_\mathrm{det}|$. 

 In order to assign a likelihood (Section~\ref{sec:likelihood} below), we require total size of the comparison sample to be sufficient such that no significant stochastic differences between distributions occur between realisations for fixed $\Theta$. We aim for a sample size of $3000$, and therefore randomly draw this many from $\tilde{\zeta}_\mathrm{det}$ to give our comparison sample $\tilde{\zeta}$. We must therefore choose the size of our initial sample $|\tilde{\zeta}_\mathrm{all}|$ such that $|\tilde{\zeta}_\mathrm{det}|>3000$ for the majority of $\Theta$. We find that $|\tilde{\zeta}_\mathrm{all}| = 1.2\times 10^5$ is nearly always sufficient. In rare cases where it is not, we draw $\tilde{\zeta}$ with the same size by allowing multiple drawings of the same synthetic planet $i$. This should not significantly change our test statistic except in cases where the fraction of detections are so low that any such $\Theta$ would in any case be highly disfavoured.

\subsubsection{Estimating likelihood}
\label{sec:likelihood}

We require a statistical test to measure the agreement of the synthetic sample of periods and radii $\tilde{\zeta}$ obtained from our model with parameters $\Theta$ with the observed data $\zeta$. \citetalias{Mulders18} estimate the likelihood using Fisher's method, by which one can combine likelihood estimates on \textit{independent} comparison quantities to estimate an overall likelihood \citep{Fisher25}. However, performing comparison tests independently for radius and period neglects any possible covariance between the two. We address this problem by performing a single statistical test quantifying the (dis-)similarity of $\tilde{\zeta}$ and ${\zeta}$ in $P_\mathrm{orb}{-}R_\mathrm{pl}$ space.

One can demonstrate differences between two two-dimensional datasets by generalising the Kolmogorov–Smirnov (KS) test. The approach for this was proposed by \citet{Peacock83}, the philosophy of which was to replace the maximal difference between cumulative distributions with the analogous difference $D$ across all quadrants in a dataset defined at each data point. While such a metric is not distribution-free, unlike the one-dimensional case, it was shown by \citet{Fasano87} that it is nearly distribution-free if the two distributions have a similar (Pearson's) correlation coefficient $\rho$. 

\citet[][see Section 14.8]{Press07} used the results of \citet{Fasano87} to infer an approximate formula for the test statistic in two-dimensions. The corresponding probability of obtaining two samples from the same distrubution is:
\begin{equation}
\label{eq:KS2D}
    p_\mathrm{KS,2D} \approx \mathcal{Q}_\mathrm{KS} \left(d \right),
\end{equation}where
\begin{equation}
    d = \frac{\sqrt{N}\bar{D}}{1+\sqrt{1-\rho^2}\left(0.25-0.75/\sqrt{N}\right)},
\end{equation}
$N$ is the effective sample size:
\begin{equation}
    N = \frac{|\tilde{\zeta}| |{\zeta}|}{|\tilde{\zeta}| +|{\zeta}|},
\end{equation}and 
\begin{equation}
    \mathcal{Q}_\mathrm{KS}(x) \equiv 1-\mathcal{P}_\mathrm{KS}(x) = 2\sum_{j=1}^\infty (-1)^{j-1} \exp \left( -2jx^2 \right)
\end{equation} is the complement of the cumulative distribution function of the KS distribution ($\mathcal{P}_\mathrm{KS}$). Since the value of $D$ can be defined at vertices drawn from either $\tilde{\zeta}$ ($D_{\tilde{\zeta} }$) or $\zeta$ ($D_{{\zeta} }$), in practice the mean $\bar{D}=(D_{\tilde{\zeta} }+D_{\zeta} )/2$ is used. Equation~\ref{eq:KS2D} is a good approximation for $N\gtrsim 20$ and $p_\mathrm{KS,2D}<0.20$. The former is no problem in this context. The latter simply indicates that two samples are not significantly different, and in our case that there are not sufficient constraints from the data to choose between model parameters. In fact, in Section~\ref{sec:MCbestfit} we do find that the upper limit model, for a finite range of $\Theta$, is not significantly different from the observed distribution. We discuss the implications in that section.

We apply the Markov chain Monte Carlo (MCMC) implementation \textsc{Emcee} \citep{ForemanMackey13} to explore the multi-dimensional parameter space. For observed data $\zeta$, Bayes' theorem states that the likelihood of a model with parameters $\Theta$ is:
\begin{equation}
    \mathcal{L}(\Theta | \zeta) \propto p(\zeta| \Theta),
\end{equation}the probability of obtaining the data from the model. In this case, the $p$-value from a two-dimensional KS test is the estimated probability of obtaining the observed data given that it comes from the same parent distribution as the synthetic data (i.e. the model). Hence $p(\zeta| \Theta) \approx p_\mathrm{KS,2D}$, as defined in equation~\ref{eq:KS2D}. Adopting uniform priors within a given range, we have:
\begin{equation}
\label{eq:logL}
    \ln \mathcal{L} = \ln p_\mathrm{KS, 2D} .
\end{equation} Equation~\ref{eq:logL} defines the likelihood we adopt for the MCMC parameter space exploration in $\Theta$.

\subsection{Results}
\label{sec:Kep_results}
\subsubsection{Best-fitting parameters}
\label{sec:MCbestfit}
\begin{table*}
    \centering
    \begin{tabular}{c | c c c c c c c c}
         Model & $\log M_\mathrm{break}/M_\oplus$ & $a_M$ &  $b_M$ & $\log M_0/M_\oplus$ & $\arctan \beta$ & $\sigma_M$ & $\log a_\mathrm{in} /1\,\rm{au}$ & $p_\mathrm{KS,2D}^{\mathrm{ML}}$ \\ \hline \hline
        \textsc{Fix-PL} & $0.93\pm0.09$ & $1.01^{+0.32}_{-0.26}$ & $-2.18^{+0.71}_{-1.72}$ & $2.12^{+0.34}_{-0.50}$ & $[\arctan 1.5]$ & $1.80^{+1.71}_{-0.76}$ & $-1.04^{+0.39}_{-0.37}$ & $0.178$ \\
         \textsc{Fix-PL-rg} & $1.13\pm0.04$ & $1.08^{+0.18}_{-0.22}$ & $-4.88^{+2.64}_{-3.19}$ & $2.22^{+0.28}_{-0.37}$ & $[\arctan 1.5]$ & $1.29^{+1.08}_{-0.54}$ & $-1.28^{+0.39}_{-0.38}$ & $0.010$ \\
         \textsc{Fix-a\_in} & $0.93\pm 0.08$ & $1.15^{+0.27}_{-0.22}$ & $-2.18^{+0.65}_{-1.40}$ & $2.52^{+0.64}_{-0.63}$ & $1.09^{+0.13}_{-0.18}$ & $1.43^{+0.72}_{-0.52}$  & $[-1.4]$ & $0.091$ \\ 
         \hline
    \end{tabular}
    \caption{Median and one-sigma range from the posterior of the MCMC parameter space exploration, and the resultant maximum likelihood $p$-value $p_\mathrm{KS,2D}^\mathrm{ML}$ for the planet mass limited models. Values with square brackets indicate that they were fixed during the MCMC exploration. }
    \label{tab:fitMCMC_iso}
\end{table*}

\begin{table*}
    \centering
    \begin{tabular}{c | c c c c c c c c}
         Model &  $P_\mathrm{break}/1\,\rm{day}$ & $a_P$ & $b_P$ & $R_\mathrm{break}/R_\oplus$ & $a_R$ & $b_R$ & $p_{\mathrm{KS,2D}}^{\rm{ML}}$\\ \hline \hline
         \textsc{BPL-PR} & $10.77^{+4.18}_{-3.13}$ & $1.72^{+0.35}_{-0.22}$ & $0.17^{+0.17}_{-0.33}$ & $3.12^{0.31}_{-0.34}$ & $0.15^{+0.26}_{-0.23}$ & $-5.84^{+2.14}_{-2.54}$ & $4.01\times 10^{-4}$ \\\hline
    \end{tabular}
    \caption{As in Table~\ref{tab:fitMCMC_iso}, but for the broken power-law period and radius model. These values can be directly compared to those obtained by \citetalias{Mulders18}, with which there is agreement across all parameters. } 
    \label{tab:fitMCMC_BRPL}
\end{table*}

\begin{figure}
    \centering
    \includegraphics[width=0.95\columnwidth]{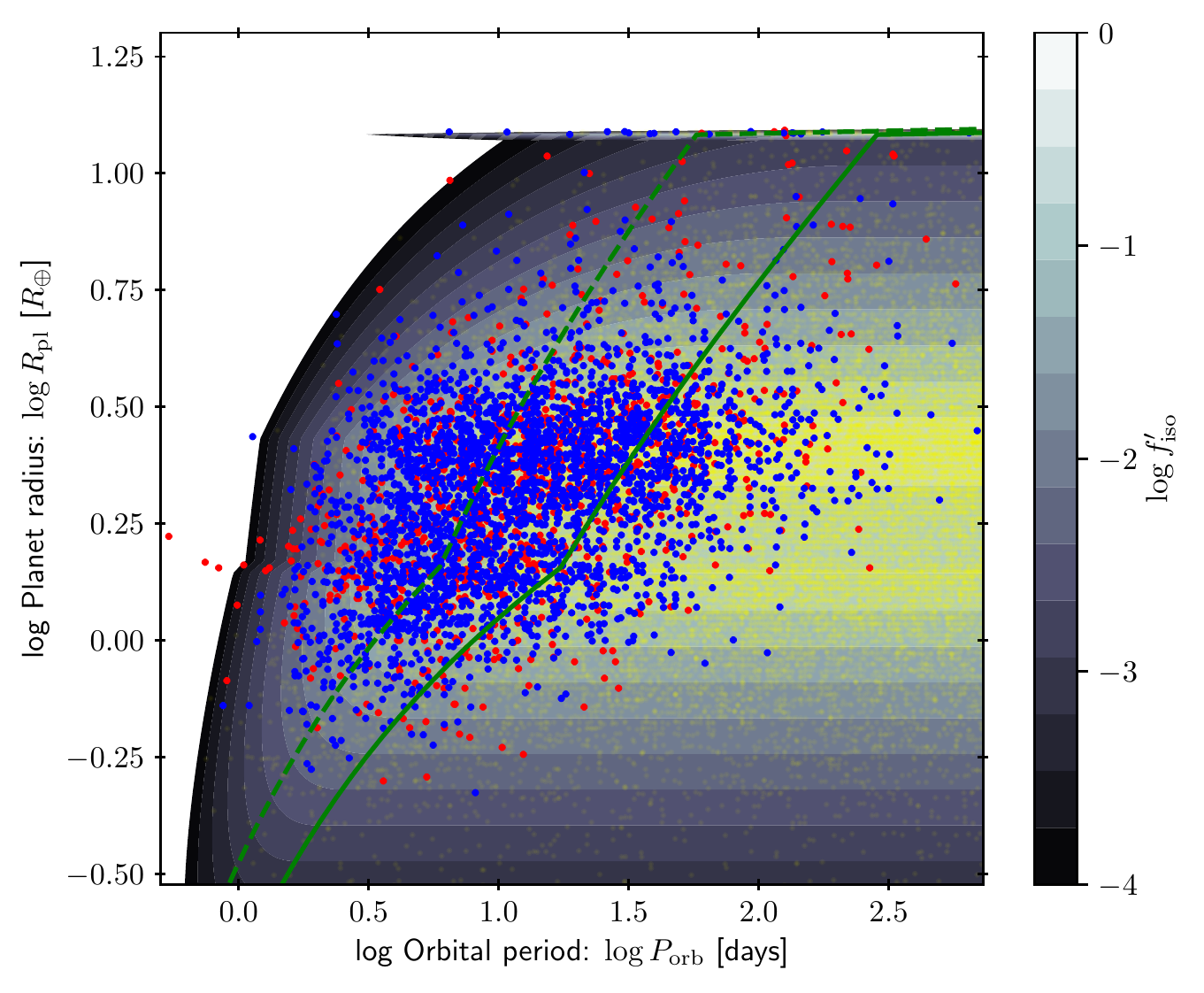}
    \caption{Example of the two-dimensional PDF in orbital period and radius resulting from the mass limited model. Red points are the observed {periods} and radii of the \textit{Kepler} multiples in our sample, faint yellow points are all planets drawn from the model PDF (the value of which is indicated by the colour bar), and blue points are those that are `detected' from this synthetic sample. The parameters are the median values from the posterior distributions obtained in the \textsc{Fix-PL} model (in Table~\ref{tab:fitMCMC_iso}), with fixed power-law index $\beta=1.5$. The solid green line is the upper limit on the planet masses (defined by mass normalisation $M_0$ and inner semi-major axis $a_\mathrm{in}$), with one deviation ($\sigma_M$) above this line indicated by the dashed green line. }
    \label{fig:PDFexample}
\end{figure}

\begin{figure}
    \centering
    \includegraphics[width=0.95\columnwidth]{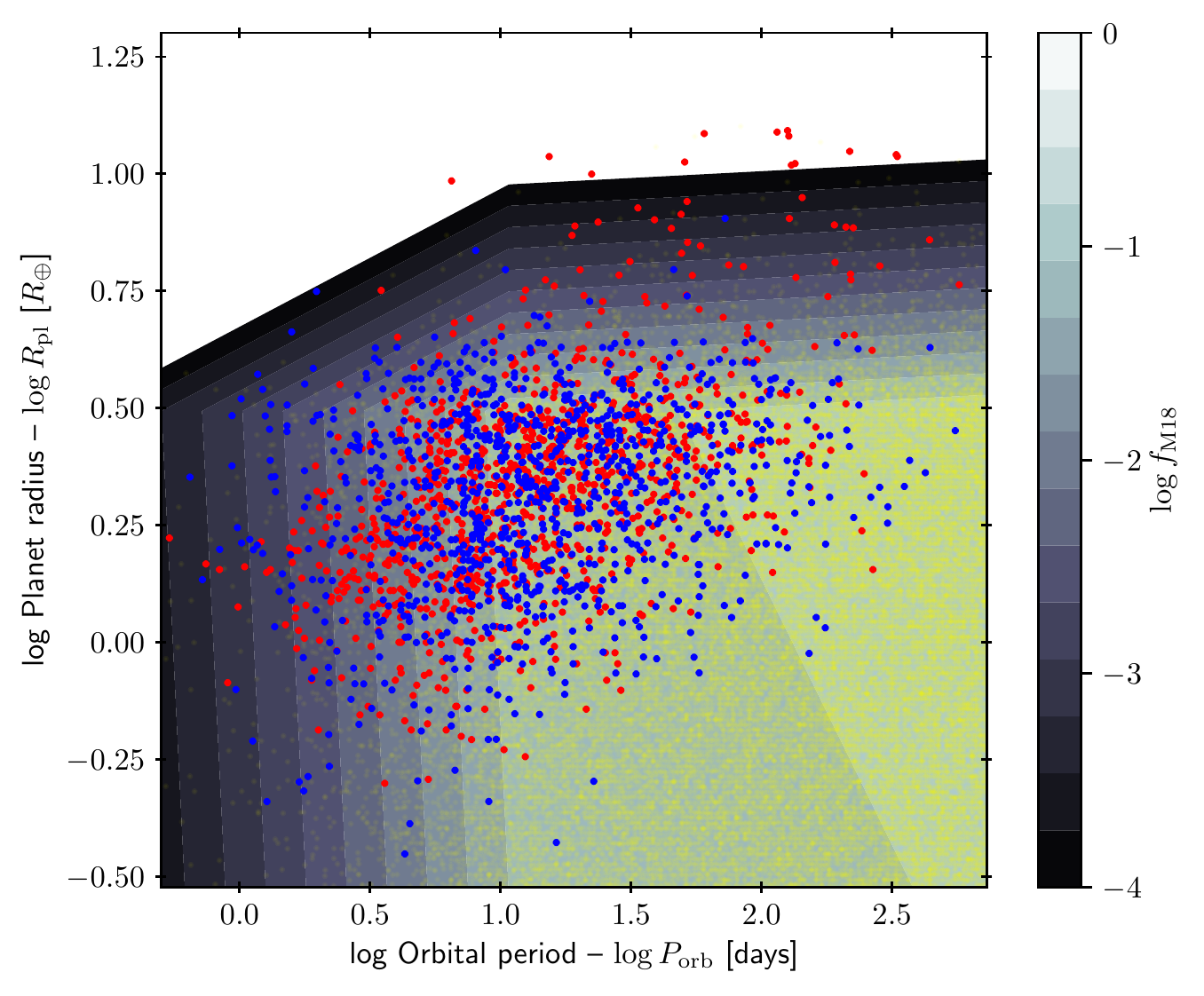}
    \caption{As in Figure~\ref{fig:PDFexample}, but for the the \textsc{BPL-PR} model, which is separable in period and radius (as in \citetalias{Mulders18}). }
    \label{fig:PDFexample_BPL-PR}
\end{figure}

\begin{figure}
    \centering
    \includegraphics[width=0.95\columnwidth]{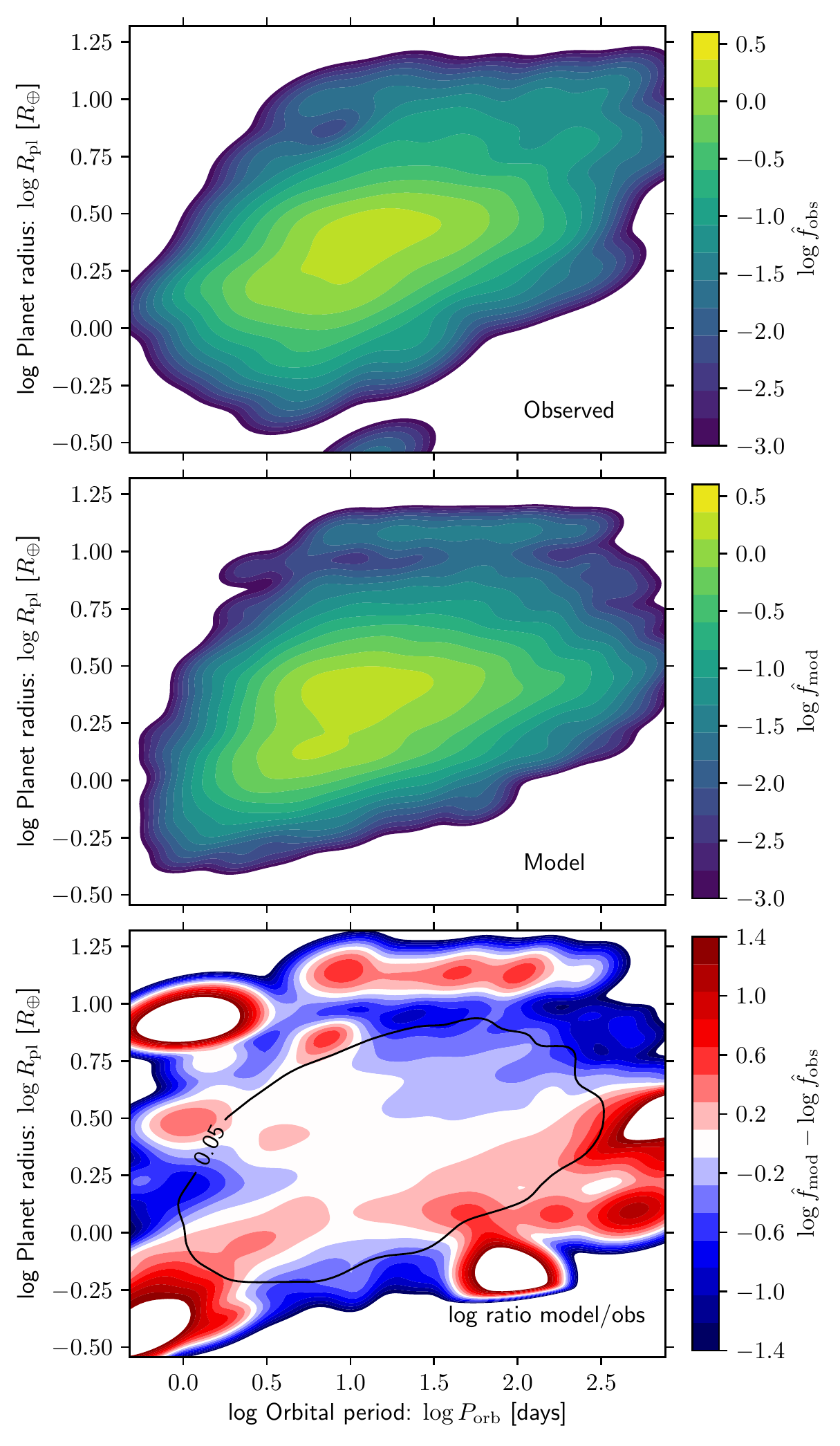}
    \caption{Gaussian KDE estimates of the two-dimensional PDFs of the observed ($\hat{f}_\mathrm{obs}$, top panel) and best fitting \textsc{Fix-PL} model with parameters in Table~\ref{tab:fitMCMC_iso} ($\hat{f}_\mathrm{mod}$, middle panel) orbital period and radius distribution of the \textit{Kepler} multiples. The bottom panel shows the (logarithmic) ratio of the model to the observed PDF ($\log \hat{f}_\mathrm{mod} -\log \hat{f}_\mathrm{obs}$), such that blue shows and underestimate and red an overestimate of the observed PDF. The black contour traces where the geometric mean $\sqrt{ \hat{f}_\mathrm{mod} \hat{f}_\mathrm{obs}} =0.05$; the region of interest is thus approximately enclosed. }
    \label{fig:residuals_fixpl}
\end{figure}

\begin{figure}
    \centering
    \includegraphics[width=0.95\columnwidth]{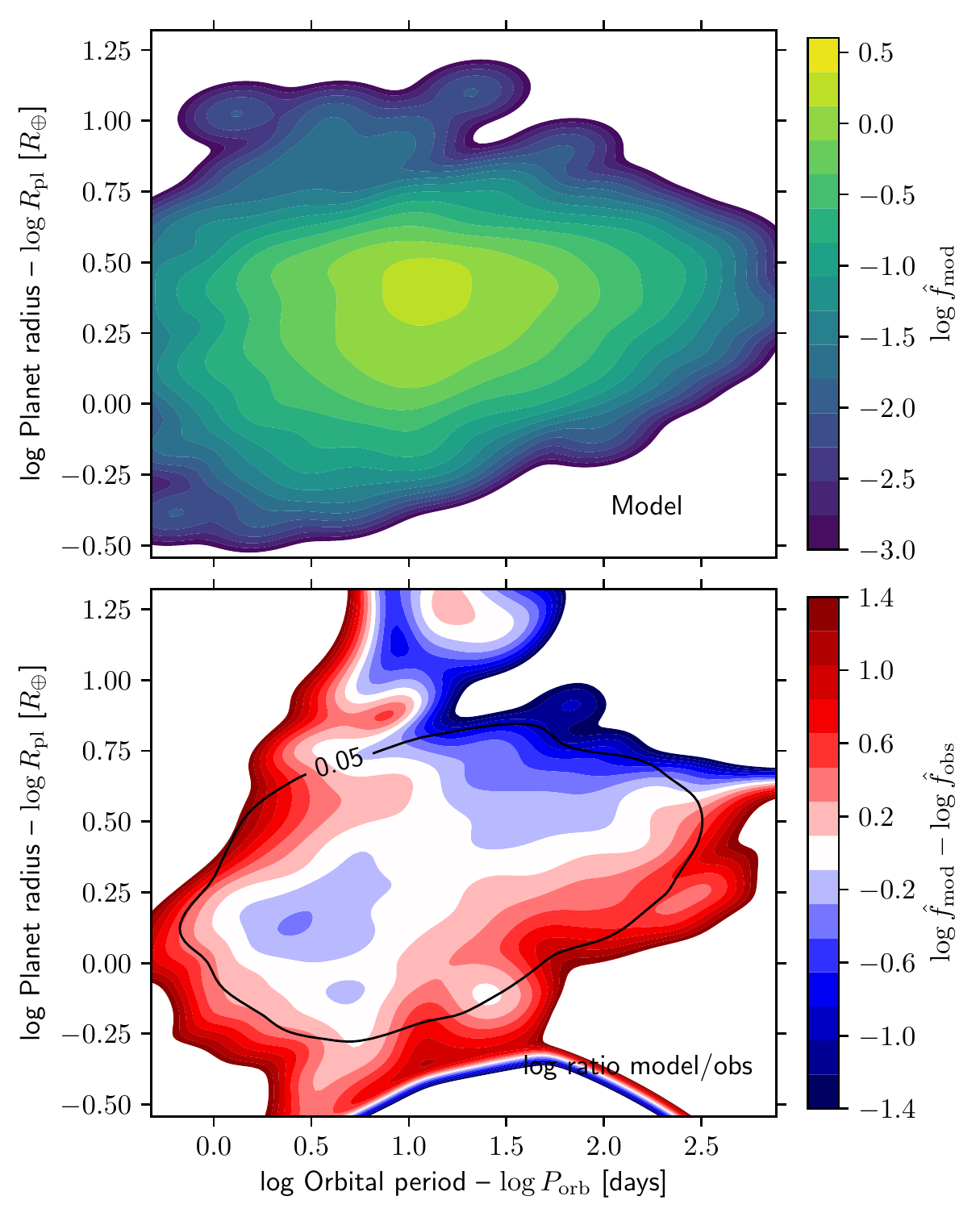}
    \caption{As in Figure~\ref{fig:residuals_fixpl} but for the \textsc{BPL-PR} model that is a separable PDF in period and radius. Such a model is similar to that of \citetalias{Mulders18}, with best-fit parameters given in Table~\ref{tab:fitMCMC_BRPL}. }
    \label{fig:residuals_brpl}
\end{figure}

The posterior distributions obtained from the MCMC exploration of our mass limited model are presented in Appendix~\ref{app:betafit}, and the results are summarised in Table~\ref{tab:fitMCMC_iso}. Initially we run a model for which we fix the power-law index $\beta=1.5$, which would be expected from the predictions of the isolation mass limited growth (Section~\ref{sec:theory_Miso}). This model (\textsc{Fix-PL}) with parameters as in Table~\ref{tab:fitMCMC_iso} is illustrated in Figure~\ref{fig:PDFexample}. Good agreement is evident across the entire parameter space, with the possible exception of at large radii $R_\mathrm{pl}\sim 12\, R_\oplus$. This is where the PDF value is strongly sensitive to the assumed mass-radius relation at high masses. For convenience, we have chosen a hard transition at $R_\mathrm{pl} =  12.1\, R_\oplus$ to a nearly flat relationship (equation~\ref{eq:MR_relation}). Smoothing this transition would yield a less abrupt increase in the PDF just above the transition radius, and thus reduced radii for a subset of the synthetic detections at the largest periods and radii. However we do not attempt to address this failing in the model because such planets account for a small fraction of the total. Indeed, by the metric of the two-dimensional KS test the observed period-radius distribution is statistically indistinguishable from that of the model, with $p_{\mathrm{KS,2D}} \approx 0.178$. 

This KS test $p$-value means that we are data-limited in determining shape of the posterior distribution rather than model-limited. This means that more restrictive constraints on the model parameters requires either alternative empirical constraints on the physical parameters (priors) or a larger sample size. It also means that adding parameters cannot improve the model by any information criterion that introduces a penalty for degrees of freedom (e.g. Bayesian/Akaike information criterion -- BIC/AIC). 

{The analogous posterior results for the alternative mass-radius relationship we discuss in Section~\ref{sec:MR_relation} (equation~\ref{eq:MR_Fulton}) are labelled \textsc{Fix-PL-rg}  in Table~\ref{tab:fitMCMC_iso}. The KS test result indicates a worse fit for the \textsc{Fix-PL-rg} model with respect to the \textsc{Fix-PL} model. This may be because our choice of mass-radius relationship in the former model is not an empirical fit. However, we are mainly interested in whether an alternative prescription for the transition between the rocky and gaseous planets alters our results. In this regard, we find marginally different best-fitting parameters describing the planet mass-function. However, the main parameters of interest are those that describe the hypothesised upper mass limit. These are consistent with the values obtained in our fiducial (\textsc{Fix-PL}) model. We conclude that the form of the inferred upper limit is not strongly dependent on the choice of mass-radius relation.  }

While a fixed $\beta=1.5$ is consistent with the observational constraints, we can ask the degree to which this index is statistically favoured. We may therefore attempt a further parameter exploration allowing $\beta$ to vary. However, given the model outcome is indistinguishable from the data for fixed $\beta=1.5$, allowing an extra parameter will always yield over-fitting and poor constraints on parameters. In particular, we find that allowing the inner edge $a_\mathrm{in}$ to vary (particularly $a_\mathrm{in}\gtrsim 0.1$~au) results in strong degeneracy between the parameters describing $f_\mathrm{lim}$, and poor constraints as a result. We therefore fix $\log a_\mathrm{in}/1\,\rm{au} = -1.4$ before repeating the MCMC exploration with $\theta\equiv \arctan \beta$ as a free parameter. This choice is semi-arbitrary: physically we expect an inner edge at a significantly greater extent than the stellar radius, and the posterior distributions are not strongly dependent on the adopted value of $a_\mathrm{in}$ for $a_\mathrm{in}\lesssim 0.1$~au. 

The results of fitting for $\beta$ are summarised in Table~\ref{tab:fitMCMC_BRPL}, with the posterior distribution shown in Appendix~\ref{app:betafit}. We first find that the parameters describing $f_M$, the broken power-law describing the planet mass PDF, are not significantly affected by allowing a variable $\beta$. We also find that the power-law and normalisation are consistent with those obtained by fixing $\beta=1.5$. The relatively poor constraints on these values are due to the strong covariance between the parameters describing the upper limit (see Figure~\ref{fig:fixain_posterior}). In particular, large $\sigma_M$ is associated with large $\beta$ and $M_0$. We are limited in that \textit{Kepler} is only sensitive to planets with short orbital periods, such that the range in period (semi-major axis) space is prohibitive. Nonetheless, the constraints in the full parameter volume space are such that further constraints on one of these parameters would allow much better constraints on the others.  

\subsubsection{Comparison to broken power-law model}

Fitting the double broken power-law model yields best-fitting parameters shown in Table~\ref{tab:fitMCMC_BRPL}, with similar results as those found by \citetalias{Mulders18}. An MC realisation for the best fitting parameters is shown in Figure~\ref{fig:PDFexample_BPL-PR}. The results of the two-dimensional KS test result for the maximum likelihood parameters yield $p_\mathrm{KS,2D}^{\mathrm{ML}} \approx 4\times 10^{-4}$, indicating a worse fit than the upper limit model. The upper limit model is therefore the preferred description of the two-dimensional distribution in $P_\mathrm{orb}{-}R_\mathrm{pl}$ space. 

In order to draw more qualitative comparisons between the two classes of model we compare the two-dimesional PDF that would be inferred if data was drawn from each model to that inferred directly from the real data. We first draw a single Monte Carlo realisation $Y_\mathrm{mod}$ of each model. We then estimate the two-dimensional PDFs from the datapoints for both model and observations $\bm{y}_i \in Y$,  where $i = 1,\dots,|Y|$ and $Y =Y_\mathrm{mod}$ or $Y_\mathrm{obs}$, the observed data. We achieve this for each dataset by using a Gaussian kernel density estimate (KDE):
\begin{equation}
    \hat{f}_Y(\bm{y})= \frac{1}{|Y|}  \sum_{i=1}^{|Y|} K_\mathrm{G} \left( \frac{\sqrt{|\bm{y}-\bm{y}_i|^\mathrm{T} C_Y^{-1} |\bm{y}-\bm{y}_i|}}{h_Y} \right),
\end{equation} where $C_Y$ is the covariance matrix and $K_\mathrm{G}$ is a Gaussian kernel with unity variance and zero mean. The shape of this PDF is dependent on the bandwidth or smoothing length scale, $h_Y$. Here we follow \citet{Scott92} in adopting:
\begin{equation}
    h_Y = |Y|^{-1/(d+4)},
\end{equation}where $d=2$ is the number of dimensions (period and radius). We apply the \textsc{Scipy} \citep{Virtanen20} implementation \texttt{gaussian\_kde} to compute the KDE here and throughout this work. 

The Gaussian KDEs for the \textit{Kepler} multiples and the fixed power-law upper limit (\textsc{Fix-PL}) model are shown in Figure~\ref{fig:residuals_fixpl}. Certain regions of $P_\mathrm{orb}-R_\mathrm{pl}$ space are poorly sampled, such that comparisons to the model are of less significance. As a guide, we indicate in the bottom panel the contour along which the geometric mean of the observed and model KDEs drops to $\sqrt{ \hat{f}_\mathrm{mod} \hat{f}_\mathrm{obs}}=0.05$. Comparisons between the model and observed distributions should be made inside the region bounded by this contour. In general we see good agreement between the two models, including a dearth of planets at large radii and short orbital periods, and a feature at the transition between gas-rich and terrestrial planets. As already suggested, the model slightly overestimates the radii of the most massive planets, which is a consequence of the adopted functional form of the mass-radius relation (equation~\ref{eq:MR_relation}). 

The \textsc{Fix-PL} model can be compared to the results of the double broken power-law model (\textsc{BPL-PR}) in Figure~\ref{fig:residuals_brpl}. In the latter case, the model significantly overestimates the number of planets with large radii ($R_\mathrm{pl}\gtrsim 4 R_\oplus$) and short orbital periods ($P_\mathrm{orb}\lesssim 10$~days) compared with the observed distribution. This can be seen immediately from the KDE, but is made clearer by comparing the bottom panels of Figures~\ref{fig:residuals_fixpl}~and~\ref{fig:residuals_brpl}. In addition, the feature at the transition from terrestrial to gaseous planets is seen as a blue blob in Figure~\ref{fig:residuals_brpl}, indicating an underestimate of the observed PDF by the model. These mismatches indicate why the broken power-law model is disfavoured by the data with respect to the upper limit model. 

\subsection{Summary}

We have demonstrated that an upper limit in planet masses can explain the distribution of the orbital periods and radii of the \textit{Kepler} multiple systems. By the metric of a two-dimensional KS test, we show that such an upper limit can yield a sample of planets that are statistically indistinguishable from the observed population. This model represents a significant improvement over a double broken power-law model describing both the period and radius distributions independently. Since the model is not distinguishable from the data, no model that invokes a greater number of free parameters can be favoured by an information criterion that introduces a penalty for degrees of freedom. Hence, this model represents the best possible description of the data in the absence of one or more of the following:
\begin{enumerate}
    \item A greater sample size (in particular, with a greater range in orbital period).
    \item Stronger constraints on physical parameters.
    \item An alternative comparison metric.
    \item A simpler model.
\end{enumerate}It is unlikely that (i) with such well characterised detection efficiencies as \textit{Kepler} will become available in the near future. More imminently, (ii) may be possible by appealing to protoplanetary disc physics and demographics, but we do not attempt to do so here. An alternative comparison metric -- (iii) -- may also be possible, although any such metric should account for the two-dimensionality of the data as we have done here. Finally, while (iv) is possible, our model is already simplified by assuming that difficult-to-treat mechanisms (e.g. migration) do not influence the planet population. 

\section{Longer period planets}
\label{sec:field_fit}

When considering whether an upper limit may also describe a subset of systems at higher mass and longer period, we are not able apply a similar approach to Section~\ref{sec:Kepmult}. This is both because of the mass-radius degeneracy for massive planets and because \textit{Kepler} planets are only confirmed if they transit three times during the four year baseline, such that a hard upper limit of two years exists in orbital period. {Although samples of $\sim 100$s of radial velocity (RV) detected planets with homogeneously analysed stellar properties exist \citep[e.g.][]{fischer05-A}}, no sample with a single discovery instrument currently compares to \textit{Kepler} in both the number of planets and the characterisation of the detection efficiency. Constraining a two-dimensional model PDF in the same way is therefore not feasible for the longer period planets. We therefore take an alternative approach. 

\label{sec:ldhd_comp}

\subsection{Motivation and sample}

{For inner planets, scattering by gravitational interaction with an exterior body most frequently results in angular momentum loss, and therefore the decrease of the semi-major axis \citep[e.g.][]{Heggie96}. Such inwards migration may result from system instability and planet-planet scattering, which may be the origin of hot Jupiters \citep[e.g.][]{Carrera19}. These dynamical instabilities may develop in an isolated system, but scattering can also be induced as a result of an external perturbation during the passage of a neighbouring star in an open cluster \citep[e.g.][and discussion in Section~\ref{sec:hem}]{shara16-A, Cai17}. If such perturbations produce a significant fraction of detected short period massive planets, and there exists a physical upper limit for planet growth as we have inferred in Section~\ref{sec:Kepmult}, we would then expect planets that form in high density environments (HDEs) to preferentially be found above it with respect to those born in low density environments (LDEs). }

Kinematic signatures of a star's formation environment may persist for $\sim$Gyr time-scales. This possibility is supported by the chemical-kinematic correlations between neighbouring stars and simulations tracing co-moving pairs of stars \citep{kamdar19a-A, kamdar19b-A, Nelson21}, as well as recent results indicating that dynamical heating is inefficient for open clusters \citep{Tarricq21}. It may therefore be possible to identify exoplanet hosts that formed and evolved in {LDEs} by their kinematics. 

\citet{Winter20c} recently delineated exoplanet hosts by position-velocity density and found significant differences between architectures of systems around hosts in the age range $1-4.5$~Gyr. Systems around lower density hosts may be associated with a higher multiplicity rate among \textit{Kepler} planets \citep{Longmore21}, hinting that these planets are less likely to have experienced dynamical perturbation. Here we compare the {HDE} and {LDE} population of planets around $\sim 1\,M_\odot$ mass stars discovered by RV surveys. We restrict the sample to those with masses $0.7{-}1.5\, M_\odot$ (as quoted in the NEA) such that they are close to $1\, M_\odot$ (within $\sim0.15$~dex). While the samples are heterogeneous such that occurrence rates cannot be directly calculated \citep[although see also][]{dai21}, the approach for delineation does not yield systematic differences in the observational biases between each sample (i.e. no bias of biases). Therefore, in principle {differences} between {HDE} and {LDE} systems can be studied (although see discussion regarding stellar ages in Section~\ref{sec:age_discuss}). To maximise detectability, we consider planets with masses $M_\mathrm{pl}\sin i>300\, M_\oplus$ leaving a sample of $68$ {HDE} and $22$ {LDE} planets. To rule out the possibility of tidal inspiral affecting our sample (see Section~\ref{sec:age_discuss}) we additionally exclude all the (massive) planets with semi-major axes $a_\mathrm{pl}>0.2$~au, yielding $61$ and $20$ planets respectively. 

\subsection{High density versus low density environment comparison}

{We wish to compare both {HDE} and {LDE} planet populations to the hypothesised limit inferred from \textit{Kepler} multiples. For any planet, we can define the logarithmic deviation from this limit:
\begin{equation}
    \delta_{M,\mathrm{Kep}} \equiv \log \frac{M_\mathrm{pl}}{M_\mathrm{lim}}
\end{equation}such that planets below the limit have $\delta_{M,\mathrm{Kep}}<0$ and planets above have $\delta_{M,\mathrm{Kep}}>0$. A caveat to this approach is that we take the results for the fixed power-law index ($\beta=1.5$) fit, since otherwise the location of the limit extrapolated to wider separations is subject to large uncertainties.  }

\begin{figure}
    \centering
    \includegraphics[width=0.95\columnwidth]{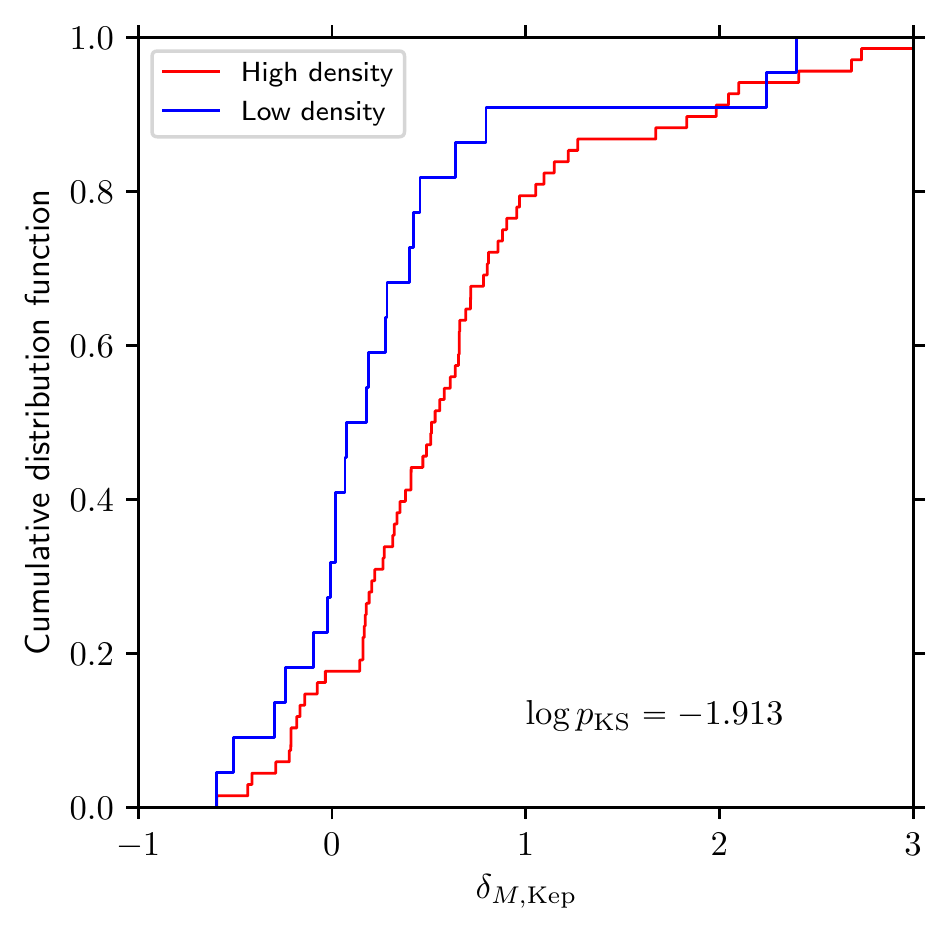}
    \caption{The distribution of the logarithmic ratio of planet mass the upper mass limit inferred from \textit{Kepler} multiples (${\delta}_{M, \rm{Kep}}$) for the {LDE} (blue lines) and {HDE} (red lines) samples with $M_\mathrm{pl}\sin i>300\, M_\oplus$. The two hot Jupiters in the {LDE} sample have outer companions. }
    \label{fig:fracabove_HJs}
\end{figure}

{We show the result of computing the ratio of planet masses to the \textit{Kepler} multiple limit in Figure~\ref{fig:fracabove_HJs}. The {LDE} planets preferentially have masses close to (or below) the extrapolated limit from the \textit{Kepler} multiples with moderate significance $p_\mathrm{KS} = 1.2\times 10^{-2}$, suggesting that this growth limit for unperturbed planets persists out to larger separations. The distribution of masses above the \textit{Kepler} limit appears as an almost continuous distribution for the {HDE} planets, while the two outlying hot Jupiters around {LDE} host stars also have outer companions (HIP 91258 b -- \citealt{moutou14-A} -- and HD 68988 b -- \citealt{Wright07}). These hot Jupiters may therefore also have reached their present orbits due to a dynamical perturbation, as suggested by the enhanced multiplicity fraction of massive hot Jupiters \citep[][]{Belokurov20-A}. Indeed, among lower mass hot Jupiters ($50 \, M_\oplus \lesssim M_\mathrm{pl}\lesssim 1 \, M_\mathrm{J}$) that are not associated with enhanced binary fractions, the RV {HDE} sample includes ten, versus none in the {LDE} sample.}

\begin{figure}
    \centering
    \includegraphics[width=0.95\columnwidth]{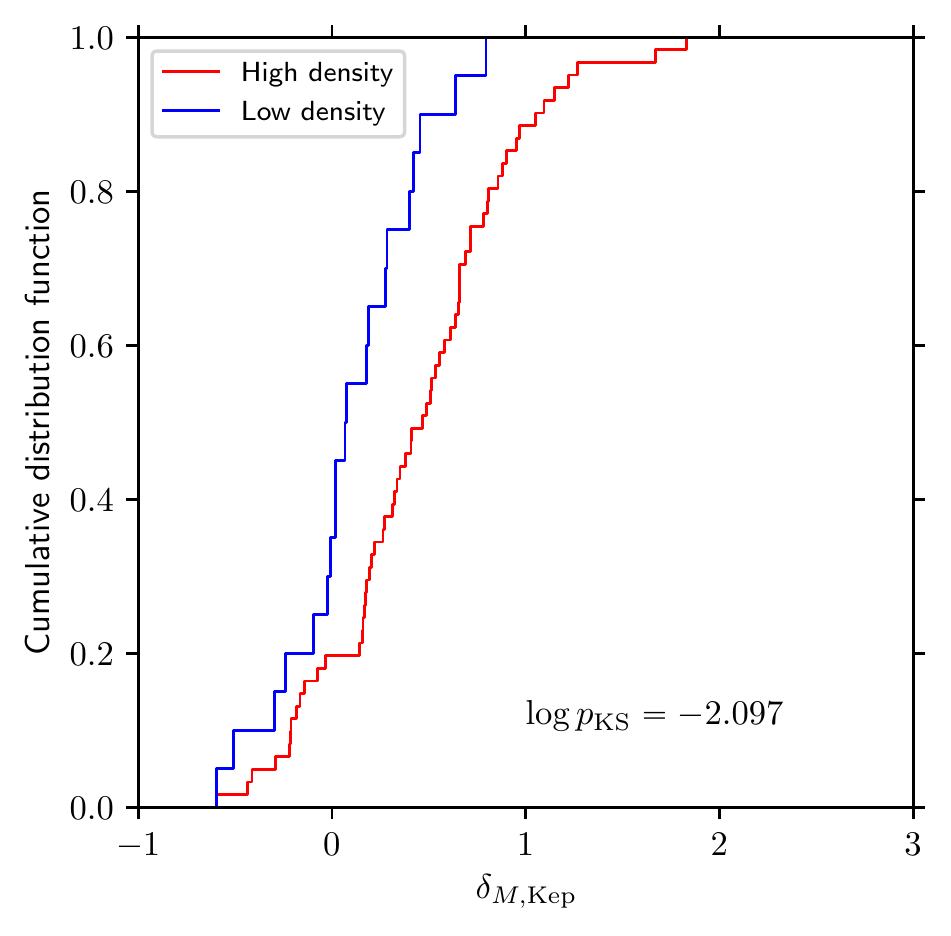}
    \caption{As in Figure~\ref{fig:fracabove_HJs} but excluding planets with $a_\mathrm{pl}<0.2$~au (hot Jupiters).  }
    \label{fig:fracabove}
\end{figure}

{To mitigate the possibility of tidal inspiral influencing our findings \citep{Hamer19}, we show the results for the sample with $a_\mathrm{pl}>0.2$~au in Figure~\ref{fig:fracabove}. We find a similar trend, with slightly greater significance $p_\mathrm{KS} \approx 8\times 10^{-3}$. This suggests that the findings are not driven purely by an increased frequency of hot Jupiters among the dense population. }

\begin{figure*}
    \centering
    \includegraphics[width=0.6\textwidth]{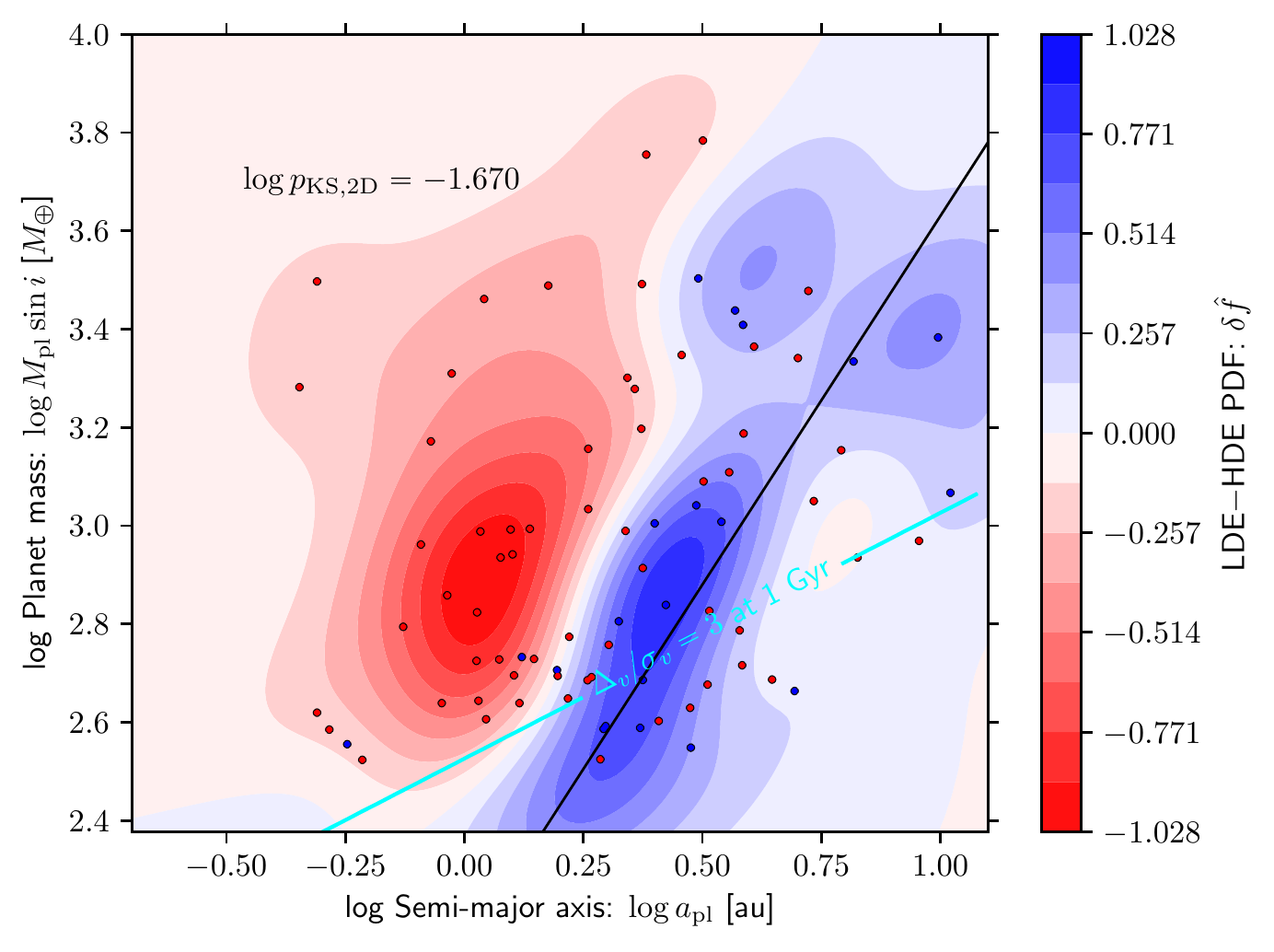}
    \caption{The distribution of planet semi-major axes ($a_\mathrm{pl}>0.2$~au) and masses ($M_\mathrm{pl}\sin i >300\, M_\oplus$), divided into {LDE} (blue) compared to {HDE} (red). The colour bar represents the result of subtracting the {HDE} KDE $\hat{f}_\mathrm{h}$ from the {LDE} KDE $\hat{f}_\mathrm{l}$, $\delta \hat{f} = \hat{f}_\mathrm{l}- \hat{f}_\mathrm{h}$. The black line is the extrapolated best-fitting upper limit inferred using the \textsc{Fix-PL} model in Section~\ref{sec:Kepmult}. The cyan line traces the contour of fixed RV variation semi-amplitude $\Delta_v$ at three times the RV jitter $\sigma_v$ due to stellar activity for a $1$~Gyr old ($\sigma_v = 10$~m~s$^{-1}$) FGK star. }
    \label{fig:pl_ud-od}
\end{figure*}

To illustrate our findings, we again consider the KDEs for our two samples. We define:
\begin{equation}
    \delta \hat{f} = \hat{f}_\mathrm{l} - \hat{f}_\mathrm{h}
\end{equation} where $ \hat{f}_\mathrm{l} $ and $ \hat{f}_\mathrm{h}$ are the Gaussian KDE in $a_\mathrm{pl}{-}M_\mathrm{pl}\sin i$ space for the planets around {LDE} and {HDE} stars respectively. The results of this calculation are shown in Figure~\ref{fig:pl_ud-od}. It is apparent that a higher fraction of the planets around stars in {LDE}s fall close to or below the upper limit inferred from the \textit{Kepler} multiples, and extrapolated out to greater separations (black line). We perform a two dimensional KS test as before, finding marginal significance. However, since we are specifically interested in the deviation from the \textit{Kepler}-inferred limit, a test across the orthogonal axes is not the best demonstration of the differences in this case.

{We do not find any significant difference in the orbital eccentricities, as defined in the NEA, between the {HDE} and {LDE} samples (both with median $e\approx0.2$). Heterogeneous determinations may influence this finding by effectively introducing noise to the distributions.}

\subsection{Possible correlations with stellar age}
\label{sec:age_discuss}

{Dynamical heating of stars yields lower phase space densities over time. Hence a sufficiently large random sample of stars with accurate stellar age determinations should exhibit differences in age when divided into high and low position-velocity density. In this case, potential covariance between stellar age and detected planet properties must be considered \citep[see discussion by][]{Mustill21}. \citet{Winter20c} used a sample that had nominal ages from the NEA between $1{-}4.5$~Gyr, such that the {HDE} and {LDE} samples did not have significantly different age distributions. However, stellar ages, even when homogeneously determined, are often challenging to accurately assess. For example, using a small set of RV host stars in a slightly larger age range ($0.5${-}$5$~Gyr),\citet{Adibekyan21} demonstrated {LDE} planets have longer orbital periods with respect to {HDE} planets (cf. Figure~\ref{fig:pl_ud-od}).  However, they also found different age distributions when adopting the theoretical isochrones by \citet{Bressan12}. In general, difficulties in accurately determining stellar ages, particularly when using metallicity or stellar kinematics that may exhibit covariance with birth environment, mean that the possibility of systematic differences as a driver of the apparent difference in exoplanet system architectures must also be considered.}

\subsubsection{Physical effects}

{Physically, one way in which age may play a role in determining exoplanet properties is due to the tidal inspiral of hot Jupiters. \citet[][see also \citealt{Mustill21}]{Hamer19} found hot Jupiters are more common for dynamically colder (high density) host stars, but concluded stellar age, and not environment, was the most likely origin for this difference. However, this finding does not persist for lower mass short period planets \citep{Hamer20}. This implies either a specifically tuned and planet mass dependent tidal dissipation efficiency, or that tidal inspiral is not the origin for the stellar kinematic dependence of hot Jupiter incidence. Arguing against tidal inspiral, hot Jupiter rates are enhanced and not depleted in the dense and old ($\sim 4.5$~Gyr) open cluster M67 \citep{Brucalassi16}.  In addition, while \citet{Adibekyan21} {are able to distinguish somewhat older ages for {LDE} stars, and fractionally fewer short period planets, they do not find differences in ages between stars hosting long and short period planets. This suggests a stronger covariance of planet architectures with dynamical properties than with (measured) age.} }

{The tidal dissipation time-scale for orbiting planets is a strong function of semi-major axis \citep{Adams06}, such that cold or warm Jupiters would not be expected to inspiral during their lifetimes. We have excluded hot Jupiters in our comparison, effectively ruling out tidal inspiral as a factor in depleting the short period planets. We recover a similar finding that massive planets on short period orbits are associated with high phase space densities. This suggests that tidal inspiral is not responsible for our findings.}

{An alternative age-related influences on a planetary system orbiting at several au would be planet-planet scattering or Kozai-Lidov oscillations \citep[e.g.][]{Naoz11}. If systems are frequently formed in a state of marginal dynamical stability, the probability of scattering may increase with the age of the system. Scattering would be expected to produce planets on short period orbits. Therefore younger (possibly HDE) planets should be found with longer orbital periods with respect to older planets. This is the opposite of our findings, and therefore not a convincing explanation for the differences between the LDE and HDE planets.} 

{We conclude that age is physically unlikely to be the origin of the differences for massive planets that are not hot Jupiters. If our findings do have a physical origin, then we might also infer that hot Jupiters are an extension of the (scattered) short period planet population. In this case, the enhanced hot Jupiter rate around dynamically colder (high density) stars \citep{Hamer19} could be due to slower dynamical heating of open clusters \citep{Tarricq21} rather than tidal inspiral. Some combination of both scattering and inspiral remains possible.}

\subsubsection{Observational effects}
{It remains possible that our results are affected by a correlation between age and the detectability of planets. Stellar activity can result in a jitter that can make detection and confirmation of true planet signals challenging. Given uncertain ages, this may influence our findings. The jitter amplitude at visible wavelengths for FGK stars of age $\sim 600$~Myr is $\sigma_v \sim 13{-}16$~m~s$^{-1}$ \citep{Paulson04, Quinn12}, and an order of magnitude greater than this for ages $\sim 100$~Myr \citep{Paulson06}. In the age range $1{-}4.5$~Gyr, $\sigma_v$ decreases comparatively slowly from $\sim 10$~m~s$^{-1}$ to $\sim 5$~m~s$^{-1}$ \citep{Hillenbrand15}. }

\begin{figure}
    \centering
    \includegraphics[width=0.95\columnwidth]{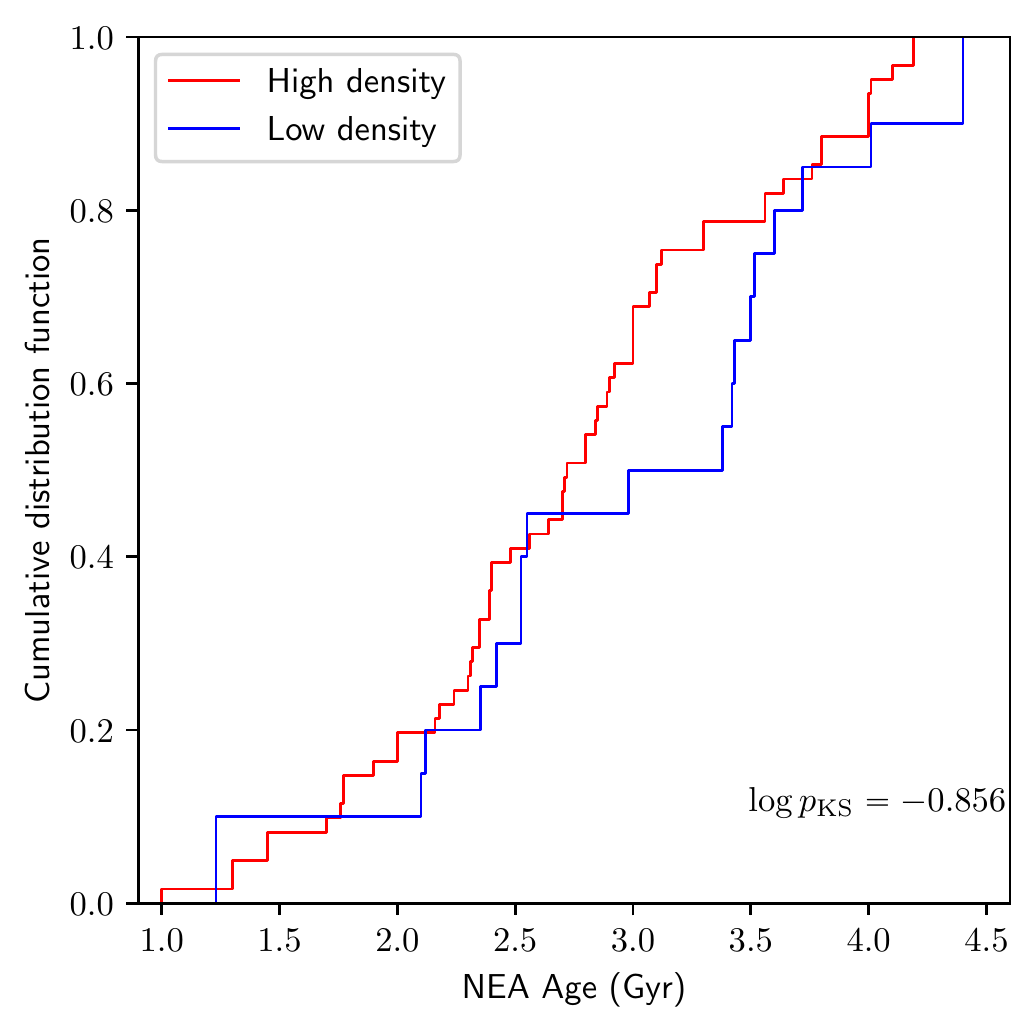}
    \caption{NEA age distribution for the exoplanet host stars of the RV detected planets with $M_\mathrm{pl}\sin i >300 \, M_\oplus$ in the {HDE} (red) and {LDE} (blue) samples.  }
    \label{fig:oldyoung_pdf}
\end{figure}

{An age-related bias is more challenging to categorically rule out, both because of the aforementioned issues determining stellar ages and because per-star completeness among RV planet hosts have not been quantified in the same way as for the \textit{Kepler} stars. However, age-related biases are unlikely to be the origin of our results for several reasons:}
\begin{itemize}
\item {The semi-amplitude of the RV signal for a planet on a circular orbit is:
\begin{equation}
    \Delta_v = 28.4 \left( \frac{M_\mathrm{pl} \sin i}{1\, M_\mathrm{J}}\right) \left(\frac{m_*}{1\,M_\odot}\right)^{-1/2} \left( \frac{a_\mathrm{pl}}{1\,\rm{au}} \right)^{-1/2} \, \rm{m \,s}^{-1}.
\end{equation}We have excluded planets with masses $<1\, M_\mathrm{J}$ such that the ratio of the semi-amplitude $ \Delta_v/\sigma_v>3$ nearly everywhere in the parameter space for stars older than $1$~Gyr (see cyan line in Figure~\ref{fig:pl_ud-od}).}

\item {The transition between {HDE} and {LDE} regimes in Figure~\ref{fig:pl_ud-od} is coincident with the \textit{Kepler} inferred limit and not coincident with contours of constant $\Delta_v$ (i.e. $M_\mathrm{pl}\propto a_\mathrm{pl}^{1/2}$).}

\item {The difference between the {HDE} and {LDE} samples appears to be an absence of easily detectable massive planets on short periods in the {LDE} sample, rather than a lack of difficult to detect low mass planets on long period orbits in the {HDE} sample (i.e. the red region in Figure~\ref{fig:pl_ud-od} lacks blue points, but the blue region does not lack red points). In the absence of a comprehensive list of RV target stars in the age range $1{-}4.5$~Gyr, we cannot directly calculate the absolute detection rates in each sample. In principle, our findings could therefore be due to a greater number of RV target stars in {HDEs} than {LDEs}. However, for a given total number of `easily detectable' planets $N_\mathrm{det}$, we can estimate the probability of $N_\mathrm{lde}$ of these detections being in the {LDE} sample, given the fraction $f_\mathrm{lde}$ of stars with a density categorisation being low density (equal numbers of {HDE} and {LDE} RV target stars corresponds to $f_\mathrm{lde}=0.5$). For equal occurrence rates, the probability of obtaining $N_\mathrm{lde}$ {LDE} planets or fewer in this domain is:
\begin{equation}
p_\mathrm{null} = \sum_{n=0}^{N_\mathrm{lde}}\begin{pmatrix}
 N_\mathrm{det} \\ n 
\end{pmatrix}f_\mathrm{lde}^{n} (1-f_\mathrm{lde})^{N_\mathrm{det}-n}.
\end{equation} In the red region of Figure~\ref{fig:pl_ud-od}, which we assume is above some age-related detectability threshold, we have $N_\mathrm{det}=36$ and $N_\mathrm{lde} = 2$. The total fraction for all stars in the solar neighbourhood varies between $f_\mathrm{ld} \approx 0.3{-}0.6$, for which equal occurrence rates can be rejected with $p_\mathrm{null} \lesssim 3 \times 10^{-4}$. Note that this value is smaller than the previously quoted null probabilities because we are here arguing from absolute numbers across samples, rather than relative numbers within each sample. Uncertainty is introduced to this test because it depends on the true $f_\mathrm{lde}$ within the superset of RV target stars. However, RV surveys preferentially select inactive stars, hence any bias should be towards older, possibly lower density target stars (increased $f_\mathrm{lde}$). Our results therefore suggest an absence of short period high mass planets around {LDE} stars, {rather than a bias-related reduction in the fraction of detections around {HDE} stars along the \textit{Kepler} limit. } }

\item {We find no significant correlation across the sample between the NEA age and the semi-major axis (Pearson $\rho=0.19$ and $p= 0.10$) or planet mass ($\rho=-0.10$, $p=0.38$). The age estimates are also similar across both samples (Figure~\ref{fig:oldyoung_pdf}), with no obvious excess at young ages in the high density population. We emphasise that these ages are heterogeneously determined, and thus may be subject to large uncertainties \citep[see discussion by][]{Adibekyan21}. }
\end{itemize}

{Despite these arguments, we do not rule out an age bias contributing to our results. Unfortunately, the sample required to do so directly is a large catalogue of RV target stars with characterised detection efficiencies and ages in the range $1{-}4.5$~Gyr, along with tens of detected planets out to several au. While this is not realistic, an alternative may be careful modelling of the differential detection efficiency with age at long periods. This is beyond the scope of this work. Nonetheless, the absence of planets around low density stars above the \textit{Kepler} upper limit is an intriguing coincidence, and the expected result if such a limit were physical.}

\section{Discussion}
\label{sec:discuss}

In this work, {we} show that \textit{Kepler} multiples are well-described by an upper limit that resembles the expected isolation mass for a low viscosity disc. {Planets around stars with low position-velocity space appear to preferentially follow this \textit{Kepler} inferred limit, with few planets found above the hypothesised limit compared to high density stars.} These findings suggest that planet growth in isolation yields isolation mass planets, and high mass planets on short orbital periods are the product of dynamical scattering. 

In this section we discuss the observational constraints from the literature that support or contradict the hypothesis that planet mass without any dynamical perturbation is limited by the local surface density during the protoplanetary disc phase. Assuming mass limited growth, we also discuss the processes that may result in the observed exoplanet architectures.

\subsection{Is the growth of inner planets limited by the isolation mass?}
\label{sec:Sigma_discuss}

\subsubsection{Disc surface density, viscosity and feeding zones}
\label{sec:discuss_massviscfeed}

We first consider whether the surface density that would yield an upper limit comparable to that obtained in Section~\ref{sec:Kepmult} is consistent with observations. For a solar mass star, the surface density we have assumed (see Section~\ref{sec:theory_Miso}) has the form:
\begin{equation}
\label{eq:sigma_conc}
    \Sigma(a) = \Sigma_0  \left( {a}/{1\,\rm{au}}\right)^{-\gamma},
\end{equation}where we take $\gamma=1$, appropriate for a radiatively heated disc with constant viscous $\alpha$  \citep{Sha73, Har98}. From the upper limit in planet masses we have inferred and equation~\ref{eq:Miso_units}, we can estimate $\Sigma_0 \approx 1.8\times 10^{3}$~g~cm$^{-2}$. The resultant disc mass is then:
\begin{align}
\nonumber
    M_\mathrm{d}& = \frac{2\pi \Sigma_0 a_0}{\epsilon_\mathrm{form}} \left(\frac{m_* }{1\, M_\odot} \right) \cdot \int_0^{R_\mathrm{out}} \mathrm{d} r \\
    \label{eq:Mdisc}
    &\approx \frac{0.038}{\epsilon_\mathrm{form}}\left(\frac{m_* }{1\, M_\odot} \right) \left(\frac{R_\mathrm{out} }{30\, \mathrm{au}} \right) \, M_\odot,
\end{align}for outer disc radius $R_\mathrm{out}$ and formation efficiency $\epsilon_\mathrm{form} \leq 1$. If $\epsilon_\mathrm{form}\sim 1$, then this is broadly consistent with masses of young protoplanetary discs \citep[e.g.][]{Eisner05, jorgensen09, Andrews13, Tychoniec20}. 

{We can further ask if the value we adopt for the gas disc surface density power-law index $\gamma=1$ is empirically supported. This value is the consequence of a constant viscous $\alpha$, with mid-plane temperature $T\propto a^{-1/2}$.} {Broadly, a $\gamma=1$ power-law out to some characteristic radius is consistent with the surface density profiles inferred from (sub-)mm continuum observations of protoplanetary discs, which trace the dust \citep{Andrews12, Andrews15}.} 
However, such constraints are somewhat complicated by the fact that we are primarily interested in the gas surface density profile, which dominates the mass budget. A variable dust-to-gas ratio, dust migration and grain growth \citep{Birnstiel10, Sellek20b}, as well as related opacity effects \citep[e.g.][]{Ros19, Ribas20} complicate the inference of the gas surface density from such measurements. In addition, detected planets are almost exclusively at semi-major axes $<10$~au, while most surface density constraints probe material at several tens of au separations from the host star that may follow a different surface density profile. Nonetheless, $\gamma=1$ is consistent with a number of observed surface densities in the dust \citep[e.g.][]{Tazzari16, Fedele17} and the gas \citep[e.g.][]{Huang16, Cle16}. {Recent resolved studies of the gas temperature profiles have also found agreement with the canonical power-law temperature profile \citep{Calahan21}.}

At separations $a_\mathrm{pl}\gtrsim 4$~au, if planets are allowed to grow to isolation as following the relationship inferred in Section~\ref{sec:Kepmult}, their masses would exceed $4\, M_\mathrm{J}$. Such planets may form from fragmentation rather than core accretion {\citep[for a recent review see][]{Kratter16}}, similar to brown dwarfs  \citep{Chabrier14, Schlaufman18}. At these distances one would no longer expect planet formation to be limited by the isolation mass, since growth to this limit would mean consuming the majority of the disc mass (see discussion in Section~\ref{sec:theory_Miso}).

\subsubsection{Disc induced migration and eccentricity}
\label{sec:disc_migrate}

{If the growth of exoplanets inside $\sim 1$ au is limited by the isolation mass, this suggests that efficient type II migration (once a gap opens in the disc) does not operate in this regime. While migration models may be able to produce observed systems in conjunction with other mechanisms \citep{Armitage02, Ida08, Ida10-A}, these models are all highly dependent on parameter choices and initial conditions \citep[e.g. disc surface density, viscosity, metallicity and the planet formation position and timescale --][]{Armitage07, Alexander09, Mordasini12}. In addition, problems persist in explaining both hot and warm Jupiters by disc induced migration, particularly tuning the required braking mechanisms to reproduce both the mass and period distribution of massive planets \citep{Hallatt20}.

While our results suggest type II migration is not the origin of high mass, close-in planets, it does not preclude the action of type I migration throughout the disc or type II at larger separations. There may be special locations within a disc that curtail migration such as dead zones \citep{Matsumura09} and ice-lines \citep{Ida08}. If migration is inefficient within the ice-line, this could be consistent with the apparent accumulation of planets close to this separation \citep{Fernandes19}. It is possible that planets that form inside or around the ice-line have their growth limited by the isolation mass, while planets that form further out at several au are still capable of significant disc-induced migration.}

\subsection{Possible counter examples}

\subsubsection{Hot Jupiters around young stars?}

If the growth of inner planets is in situ, and limited by the isolation mass, it might be expected that hot Jupiters around young stars should be rare. It is challenging to (unambiguously) detect planets around the youngest stars in the plane of a disc by the light curve during transit. Detection and characterisation using RVs is similarly challenging since magnetic activity can induce signals that are orders of magnitude greater than line of sight velocity variations. This means that, even once detected, constraints on planetary properties are often weak. 

Several studies have claimed the discovery of exoplanets in open clusters and associations \citep{Quinn12,Quinn14, malavolta16-A, Mann16, mann17, Rizzuto20}. The existence of hot Jupiters in open clusters has been used to suggest that hot Jupiters either form in situ or migrate though planet-disc interaction. However, hot Jupiters found in these environments are older than $\sim 600$~Myr such that tidal circularisation is possible. RV surveys of younger open clusters ($\lesssim 100$~Myr) have so far yielded no hot Jupiter discoveries \citep{Paulson06, Bailey18, Takarada20}. 

For individual very young (few Myr old) stars, follow ups on hot Jupiter candidates have cast doubt on a number of detections. One of the greatest challenges to the upper limit hypothesis we have explored in this work would have been the hot Jupiter detected around the $2$~Myr CI Tau \citep{Johns-Krull16}. This detection would be particularly problematic due to the extended protoplanetary disc hosting an ensemble of gas giants out to $\sim 100$~au \citep{Clarke18}. It would therefore be difficult to explain the hot Jupiter as a product of a dynamical perturbation. However, \citet{Donati20} presented observations constraining the magnetic activity of CI Tau, showing that the rotation period is equal to that of the hypothesised hot Jupiter ($9$~d), and that the field is sufficient to produce a magnetospheric gap extending to close to the corotation resonance. This casts doubt on the origin of the previous detection. Similarly, \citet{Damasso20} were recently unable to detect the previously claimed hot Jupiter around the $\sim 2$~Myr V830 Tau \citep{Donati16, Donati17} in HARPS-N data, despite performing a wide range of tests.  

{There remain two examples of hot Jupiter detections around weak-lined T-Tauri stars that have not yet been called into question: one around the $17$~Myr old TAP 26 \citep{Yu17} and one around the $5{-}10$~Myr K2-33 \citep{David16}. TAP 26b is detected by radial velocity variations, requiring careful modelling to infer that differential stellar rotation cannot be the origin for the signal. The period is $\sim 10$~days, but could not be uniquely determined from the data, and the minimum mass is $1.66\pm0.31\,M_\mathrm{J}$. K2-33b is inferred by transits of period $5.4$~days, and has a radius of $5.8\pm0.6\,R_\oplus$. Only upper limit constraints exist on the mass this young planet, and the radius is marginally consistent with the upper limit we infer in Section~\ref{sec:Kepmult}. For neither planet does the available data offer (strong) constraints on the orbital eccentricity. If real, TAP 26b in particular is a challenge to any formation model with no efficient type II migration through the disc. However, since no substantial disc remains and there is currently no suggestion of a stable system of exterior planets, in principle migration through the disc initiated by scattering remains possible. If dynamical scattering occurs during the disc phase, the subsequent orbital evolution of a massive planet can be unpredictable, and one must then consider how growth, migration and planet-disc interactions alter the orbit of such a planet \citep[e.g.][]{Papaloizou01, Bitsch13, ragusa18-A}.}

\subsubsection{ALMA `planets' in discs}

{Studies of dust in protoplanetary discs have demonstrated that structure is common \citep[e.g.][]{HLTau_15, Andrews18, Clarke18}. Many of these structures are dips in the dust surface density in rings that may be consistent with the clearing of material by planets \citep[e.g.][]{Zhang18}. Although other explanations for these rings are possible \citep{Zhang15,Flock15, Gonzalez15, Okuzumi16}, many discs also exhibit kinematic deviations from Keplerian velocity that are consistent with presence of protoplanets \citep{Casassus19, Pinte20}. 
 
 The best constraints on planet growth and evolution in a disc come from the two accreting protoplanets inside the transition disc around PDS 70 \citep{Hashimoto12,Keppler18, Haffert19}. The resonant configuration and accretion rates of these planets can be explained by planet-disc interaction and migration \citep{Toci20}. PDS 70 b and c have orbital separations of $\sim 20$~au and $35$~au and masses of $\sim 4{-} 17\,M_\mathrm{J}$ and $\sim 2{-}10\,M_\mathrm{J}$ respectively \citep{Muller18,Haffert19,Christaens19, Bae19}. Such planets (or sub-stellar objects) may be massive enough to have formed by fragmentation rather than core accretion \citep{Schlaufman18}, and outside of the range in which planets can possibly grow to the isolation mass without consuming the majority of the disc mass (if the surface density follows equation~\ref{eq:sigma_conc} with $\Sigma_0 \approx 1800\, \rm{g}\, \rm{cm}^{-2}$). {While it may be possible to constrain planet masses from the geometry of the dust gaps \citep{Rosotti16}, practically all gaps that have been resolved in discs are at $\gg 10$~au from the host star. It is therefore not presently possible to constrain masses of forming planets in the inner disc.} Recent developments in the processing of interferometric observations may in future be applied to resolve structures in discs down to sub-beam scales to test the hypothesis we have presented here \citep{Jennings20, Jennings21}. }

\subsubsection{Multi-planet systems in mean motion resonances}

{Mean motion resonances (MMRs) in planetary systems are thought to result from migration within a gas disc \citep[e.g.][]{Cresswell06}. {Systems in MMR are unlikely to form as a result of rapid or high eccentricity migration \citep{Quillen06, Pan17}. Multi-planet systems rarely occupy resonant configurations \citep{Fabrycky14}, which suggests that any disc migration does not ubiquitously result in resonant capture \citep[see also:][]{Hansen12, Petrovich13, Chatterjee14}.} However, there exist a number of examples of planetary systems that are in resonant configurations, such as those around the M Dwarf GJ 876 \citep[][]{Batygin15,Nelson16, Cimerman18}, the solar mass Kepler-36 \citep{Carter12}, and the very low-mass red dwarf TRAPPIST-1 \citep{Gillon17, Luger17}. }

{The existence of any resonant systems requires a degree of slow, low-eccentricity migration in some systems. In support of such migration being frequent, resonant chains have also been shown to be easily broken during type I migration \citep{Goldreich14,Hands18}. Further, \citet{Longmore21} suggest that resonant systems may be correlated with low density stellar environments, in which case MMRs resulting from isolated planet formation processes in a disc may be more frequent than previously thought. Hence, it is possible that (nearly) all planets experience slow, low eccentricity migration in the disc phase, with a subset escaping MMR configurations or undergoing subsequent dynamical perturbation. By contrast, \citet{Morrison20} demonstrate planets can form in resonant chains during oligarchic growth in a depleted gas disc, without much migration. Either scenario might give rise to a fraction of systems in MMR, while not the majority. In either case, the upper limit we suggest in this work only requires that rapid type II migration does not result in massive planets on short period orbits, and does not preclude some degree of slow type I migration. }

\subsubsection{Circumbinary planets}

{Circumbinary planets around binaries with periods $\gtrsim 10$~days have been found to be abundant, {with incidence rates comparable to planets around single stars} \citep[][although one must also factor in the high probability of transit --  \citealt{Martin15}]{Armstrong14}. {There are now 14 known transiting circumbinary planets \citep{Kostov20}, and most of them orbit very close to the innermost stable orbit around the binary. In situ formation is very unlikely \citep[e.g.,][]{Paardekooper12}, and high-eccentricity migration is impossible (due to high-eccentricity orbits being unstable). The orbits of these planets are consistent with formation at larger radius followed by disc-driven migration, as predicted prior to their discovery \citep{Pierens08}. However, the torques exerted on these planets mean that migration in such systems does not reflect the scenario in single systems, and would not necessarily contradict the hypothesised mass limit in this work.}}

\subsection{Mechanisms operating on perturbed systems}
\label{sec:mech_discuss}

We have illustrated that there may exist an upper limit for the growth of inner planets. This begs the question: what mechanisms could yield deviation from this mass in the observed exoplanet sample?

\subsubsection{Binary/multiple interactions and stellar encounters }
\label{sec:hem}

High eccentricity migration, the result of dynamical perturbation as opposed to disc migration, is a promising mechanism to produce hot Jupiters and other high eccentricity massive planets. Since a high fraction of stars are multiples \citep[][although not necessarily most -- \citealt{Lada06}]{Duquennoy91, Raghavan10}, 
one might expect some fraction of systems to be influenced by a companion. In particular, the exchange of angular momentum during Kozai-Lidov (KL) oscillations can produce extreme eccentricities and inclinations \citep{Nagasawa08,Naoz12, Naoz16}. This mechanism may produce hot Jupiters when coupled with tidal circularisation (Section~\ref{sec:tidal_circ}). Although the absence of high eccentricity transit detections in \textit{Kepler} data marginally opposes this scenario \citep{Dawson15}, this may be explained by non-steady state flux of planets inwards (early migration) or the considerable uncertainties in the circularisation time-scale. 

{Investigations into the occurrence rates of hot Jupiters around binary components are often sub-divided into short and long period binaries. \citet{Ngo16} find that hot Jupiters are associated with double the binary fraction in the separation range $50{-}2000$~au than field stars \citep[see also][]{Fontanive19-A}. The opposite trend for binaries with separations $1{-}50$~au (suppressed by a factor $\sim 4$ for hot Jupiter hosts) suggests a transition from formation and inward scattering to formation suppression. \citet{Moe19} show that, accounting for selection biases, the correlation of hot Jupiter occurrence with wider binaries only apply to the most massive hot Jupiters that did not necessarily form by core accretion. However, this is not directly comparable to the similar result of \citet{Belokurov20-A}, who find a higher frequency of tight binaries (separations $\sim 0.1{-}10$~au) among hot with respect to cold Jupiters with mass  $>1\, M_\mathrm{J}$ (the majority with masses $1{-}4\,M_\mathrm{J}$). Whether the $4\,M_\mathrm{J}$ threshold, motivated by the metallicity occurrence rate correlation \citep{Schlaufman18}, is directly related to a correlation with binary occurrence remains an open question. It is possible that lower mass hot Jupiters form at smaller separations and require a tighter binary companion to perturb them, which might reconcile the findings of \citet{Belokurov20-A} and \citet{Ngo16}. While the nature of these correlations requires further analysis, it is clear that some fraction of massive planets on short period orbits have experienced perturbation from a binary companion. }

Apart from long time-scale oscillations, rapid planet-planet scattering has been proposed as another mechanism to produce high eccentricity (massive) planets \citep[e.g.][]{chatterjee08-A, Carrera19}. The close passage of two stars can also gravitationally perturb a (proto)planetary system, inducing instabilities and scattering bodies away from their initial orbital configuration \citep[e.g. \citealt{malavolta16-A, pfalzner18-A, Cai17, cai18} -- see][for a review]{Davies2014-A}. Although such close interactions during the protoplanetary stage (within $\sim 3$~Myr) occur mainly within small-scale bound stellar multiple systems \citep{bate18-A, Win18c}, if the whole cluster remains bound over $\gtrsim 100$~Myr then planetary scattering by neighbouring stars is possible at stellar densities $\gtrsim 10^{2-3}$~stars/pc$^{3}$ \citep[e.g.][]{Malmberg07,malmberg11-A, Spurzem09,Li19-A, Stock20, Wang20, Wang20b}. Implementing N-body calculations of systems with multiple massive planets, \citet{shara16-A} find that hot Jupiter formation by scattering occurs for $\sim 1$~percent of systems at stellar densities typical of known open clusters. Scattering may also be significantly enhanced when accounting for the increased cross-sections of binaries \citep{Li15,Li20}, possibly in conjunction with KL oscillations or von Zeipel-Kozai-Lidov if the mass of inner and outer companion are comparable \citep{Naoz11, Rodet21}. Hot Jupiter formation by encounters and subsequent scattering is directly supported by the enhanced occurrence rates in the dense open cluster M67 \citep{Brucalassi16} and the younger, less dense Praesepe \citep{Quinn12}. Indirectly, multiple studies have found that hot Jupiters are more frequent at high position-velocity density \citep{Hamer19, Winter20c, Adibekyan21}. This finding may be related to the slower dynamical heating of open clusters \citep[][]{Tarricq21}, although if stellar age estimates are sufficiently uncertain the findings could also be attributed to tidal inspiral of older hot Jupiters \citep{Hamer19}. However, tidal inspiral would not explain the preference for planets in the {LDE sample} at or below the \textit{Kepler} inferred growth limit among the non-hot Jupiter planets, discussed in Section~\ref{sec:ldhd_comp} (see Figure~\ref{fig:pl_ud-od}). 

Stellar scattering preferentially influences planets at wide separations, as encounter rate (scattering cross-section) scales with $a_\mathrm{pl}$ for gravitationally focused encounters, or $a_\mathrm{pl}^2$ for hyperbolic ones. Assuming the planet remains bound post-encounter, it most frequently loses angular momentum, resulting in decreased semi-major axis \citep{Hills75,Hut83,Heggie96}. We therefore expect encounters to produce a population of planets with masses preferentially above any limit for formation in isolation. Our findings in Section~\ref{sec:field_fit} are broadly consistent with this expectation.

\subsubsection{Tidal circularisation and atmospheric photoevaporation}
\label{sec:tidal_circ}
Once a massive planet has lost angular momentum by a stellar encounter, the induced high eccentricity can result in numerous close passages with the host star. Further angular momentum loss, this time through tidal interactions with the host star, can then lead to orbital circularisation, inclination damping and shortening of the orbital period \citep{Hut81,Rasio96_tidaldecay,Ivanov04,Ivanov07, Faber05, jackson08}. Large quantities of energy can be lost during this tidal interaction, making it a promising mechanism for the formation of hot Jupiters.  Once on short period orbits, massive planets may lose mass due to irradiation and photoevaporative heating by the central star \citep{Owen12}. This photoevaporation has been suggested as the origin of the observed `radius valley' \citep{Owen17, Mordasini20}.

\subsubsection{External photoevaporation}
\label{sec:extphot}

Protoplanetary discs can be photoevaporated by massive neighbouring stars in young (few Myr old) star forming environments \citep{johnstone98-A, Ada04, Win18b}. This is both predicted theoretically and found empirically to reduce the disc mass \citep{Mau16, Ans17,Eis18,facchini16-A, Haw18,Winter19b,vanTerwisga19-A} and shorten the disc life-times \citep{Stolte04,sto10,Stolte15,fang12-A, guarcello16-A, concharamirez19-A, Winter19,Winter20}. In terms of the planet masses, one would therefore expect to reduce masses of planets that form, by some combination of reducing the local disc surface density and limiting the time-scale for formation. 

While the principles of photoevaporation are well-known, and there is now ample direct evidence to suggest it operates on discs \citep[e.g.][]{Ode94, Kim16, Haworth20}, how this may influence exoplanet architectures remains uncertain. A reasonable assumption is that photoevaporative winds should decrease the mass-budget available for planet formation, and do so preferentially at large separations from the host star \citep[][]{johnstone98-A}. {This need not necessarily suppress planet formation; photoevaporation of the gas in the disc may also result in enhanced dust-to-gas ratio, which can make the growth of planets by dust instabilities (e.g. streaming) more likely \citep{Throop05}.} Nonetheless, if discs are externally depleted in dense environments, and the most massive planets must grow from gas in the outer disc, this might explain the dearth of massive planets and brown dwarfs in globular clusters \citep{Gilliland00, Bonnell03_BDs}. 

\subsubsection{Other mechanisms}

Numerous other mechanisms may exist that drive deviation from the hypothesised mass limit. The deposition of short lived radionuclides into the discs due to supernovae or the winds of massive stars \citep[e.g.][]{Martin09} can lead to the heating and dehydration of planets \citep{lichtenberg19-A}. Ionisation by cosmic rays that may also lead to more compact discs in some star forming regions \citep{Kuffmeier20}. Enhanced metallicity in some environments may speed up planet formation and slow down disc dispersal \citep{Ercolano10}. This list is unlikely to be exhaustive.

\subsection{Ramifications for the Solar System}
\label{sec:solar_system}

With the exception of Jupiter, the planets hosted by the Sun are much lower mass than the upper limit for which we have presented evidence in this work. This may be the consequence of premature disc dispersal. The idea that the Solar System has been sculpted by its star formation environment is not a new one \citep[see e.g.][for a review]{adams10}. Evidence ranges from the the abundance of daughter isotopes of short lived radionuclides that may have originated in a supernova nearby the young solar nebula \citep[e.g.][]{Meyer00,Goswami00,Adams01,Wadhwa07, Gounelle09, Banerjee16, Boss17}, to the high eccentricities and inclinations of the trans-Neptunian bodies that could originate from a stellar encounter in a moderate density cluster  \citep{Becker18,Pfalzner18,Moore20,batygin20}. By the definition of \citet{Winter20c} the Sun occupies a high phase space density. It is possible that the Sun originated in M67, members of which have similar ages and metallicities \citep{Yadav08, Heiter14}. If such a cluster initially had $\sim 10^4$ members \citep{Hurley05}, this would be consistent with constraints from the Solar System properties \citep{portegies09, adams10, Pfalzner13, Moore20}. However, this scenario remains a subject of debate \citep{Jorgensen20, Webb20}. 

{Since the giant planets in the Solar System do not have high eccentricities, they are unlikely to have been significantly dynamically perturbed. However, if the young Solar System was exposed to strong UV fields in its birth environment, this may have reduced the gas reservoir from which planets could accrete before they were able to grow to the isolation mass. In this case, the feeding zones could remain small such that the young planets would not merge, and we may expect much lower planet masses than the limit we infer in this work. This is a speculative scenario, but a feasible way to produce a system of low eccentricity planets well below any hypothesised mass limit in the growth of planets forming in isolation.}

\section{Conclusions}
\label{sec:concs}

In this work, we investigate the distribution of exoplanet architectures in light of evidence that some subset of planetary systems have been perturbed from their formation configuration. In particular, we consider whether there may exist an upper limit in exoplanet masses that is disguised by the perturbed planets. 

For close-in planets ($\lesssim 1$~au), we present an approach for modelling the period-radius distribution of \textit{Kepler} multiples, accounting for the two-dimensionality of the data. Around solar mass hosts, a model incorporating an upper limit in the planet masses of the form:
    \begin{equation}
        M_\mathrm{lim} \approx M_0 \exp \left(- \frac{a_{\rm{in}}}{a_\mathrm{pl}}\right) \left( \frac{a_\mathrm{pl}}{1 \, \rm{au}}\right)^{\beta} \, M_\oplus ,
    \end{equation} 
can reproduce the observed distribution of periods and radii when coupled with an empirical mass-radius relation. The power-law index $\beta=1.5$ is consistent with the data, as would be expected for in situ growth limited by the isolation mass. Imposing this upper limit provides a better fit to the data than a separable PDF in period and radius. 

For longer period RV detected planets with $a_\mathrm{pl}>0.2$~au and $M_\mathrm{pl} \sin i>300\,M_\oplus$, those hosted by stars at low position-velocity density have masses that are preferentially below the \textit{Kepler} inferred limit with $\beta=1.5$. This finding cannot originate from tidal inspiral of hot Jupiters ($a_\mathrm{pl}<0.2$~au) and is not easily explained by age-related observational biases. This suggests that planetary systems around high density host stars are more likely to have been scattered inwards, consistent with the suggestions of higher planet multiplicity among host stars in low density environments \citep{dai21, Longmore21}. 

We conclude that, given the mounting evidence that a subset of planetary systems have undergone dynamical perturbation, attempting to identify planets that have not been perturbed can offer insight into the formation process.

\section*{SOFTWARE}
We thank the contributors for making the following packages, of which we made use, public: \textsc{Python}, \textsc{Matplotlib} \citep{Hunter07}, \textsc{Numpy} \citep{Harris20}, \textsc{Scipy} \citep{Virtanen20}, \textsc{Pandas} \citep{Reback20}, \textsc{Emcee} \citep{ForemanMackey13}, \textsc{KeplerPORTs} \citep{Burke17}, \textsc{EPOS} \citep{Mulders18}.

\section*{Acknowledgements}
The authors thank the referee for a positive and considered report, and Megan Ansdell, Cathie Clarke, Kees Dullemond, Sarah Jeffreson and Andrew Mann for their helpful suggestions. AJW acknowledges funding from an Alexander von Humboldt Stiftung Postdoctoral Research Fellowship. This project has received funding from the European Research Council (ERC) under the European Union’s Horizon 2020 research and innovation programme (grant agreement No 681601).

\section*{DATA AVAILABILITY}

All data used in this work has been obtained from publicly available repositories.




\bibliographystyle{mnras}
\bibliography{truncation} 


\appendix
\section{Model posterior distributions for \textit{Kepler} multiples}
\label{app:betafit}

We present posterior distributions for the models discussed in Section~\ref{sec:Kepmult}. The MCMC parameter space exploration is performed using \texttt{emcee} \citep{ForemanMackey13} with $200$ walkers and a total of $2000$ iterations, the first $500$ discarded as the `burn-in' stage. In all cases we assume uniform priors within reasonable ranges. 

\subsection{Fixed power-law index (\textsc{Fix-PL}) }

\label{app:fixpl}

\begin{figure}
    \centering
    \includegraphics[width=0.95\columnwidth]{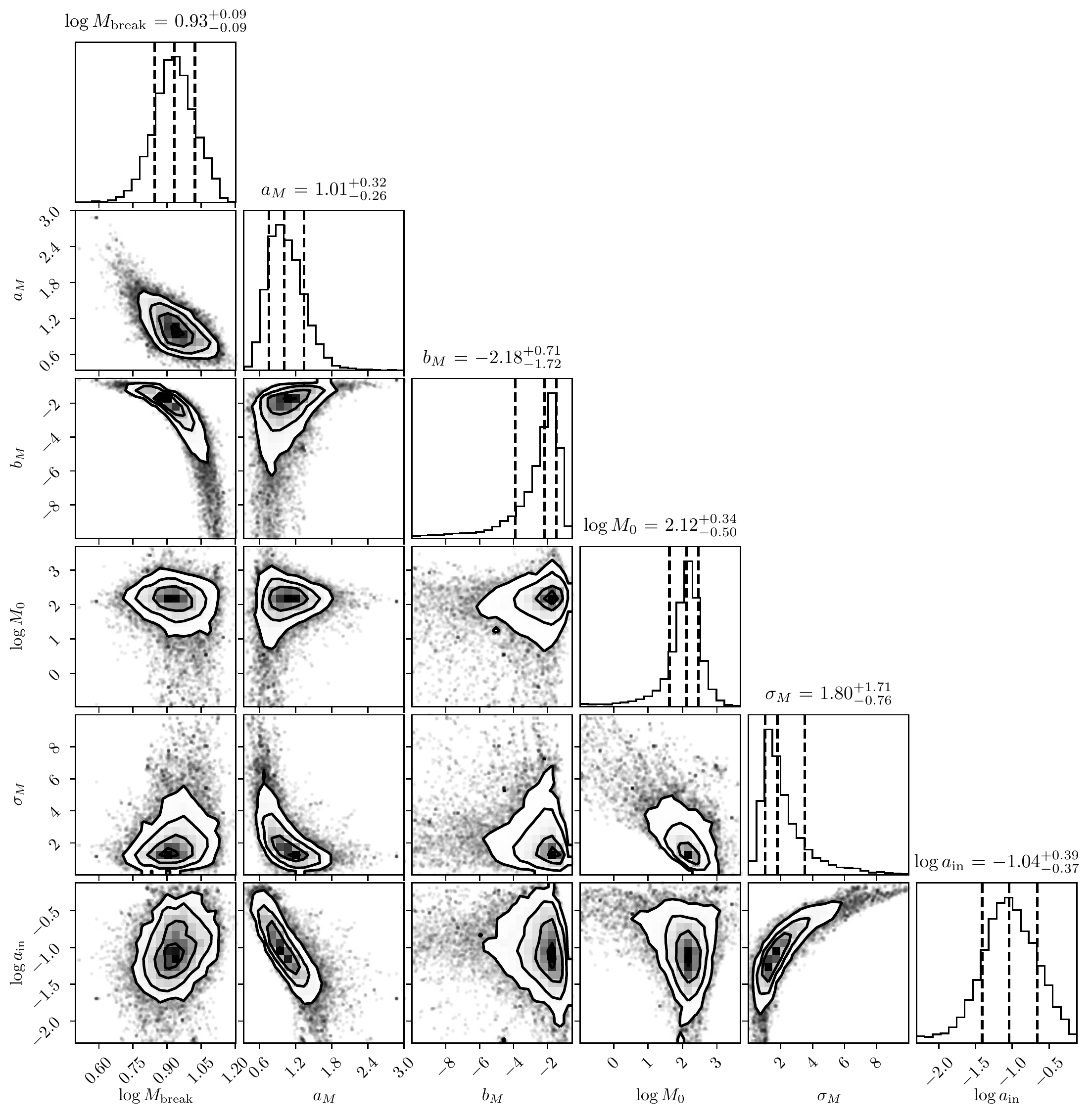}
    \caption{Posterior distributions resulting from the MCMC parameter space exploration for the fiducial mass limited model (\textsc{Fix-PL}). The quoted values for break mass $M_{\rm{break}}$ and upper limit normalisation $M_0$ are in units of $M_\oplus$, and the inner edge $a_\mathrm{in}$ is in astronomical units.  }
    \label{fig:fixpl_posterior}
\end{figure}

The posterior distribution for the upper limit model \textsc{Fix-PL}, wherein the power-law index describing the upper limit ($M_\mathrm{lim}\propto a_\mathrm{pl}^\beta$) is fixed, $\beta=1.5$, is shown in Figure~\ref{fig:fixpl_posterior}. The ranges for the priors for the mass-only component of the PDF are $-10$ to $10$ in $a_M$, $b_M$, and $1-100\,M_\oplus$ in $M_\mathrm{break}$. In the upper limit, we allow $M_0$ in the $0.1{-}5000\, M_\oplus$, $\sigma_M$ between $0$ and $10$ and $a_\mathrm{in}$ between $1\, R_\odot$ and $1$~au. We find particularly strong covariances with the value of inner-radius $a_\mathrm{in}$. Large $a_\mathrm{in}$ correlates with high $\sigma_M$, which in turn yields poor constraints on the other parameters.

\subsection{Alternative mass-radius relation (\textsc{Fix-PL-rg})}

\begin{figure}
    \centering
    \includegraphics[width=0.95\columnwidth]{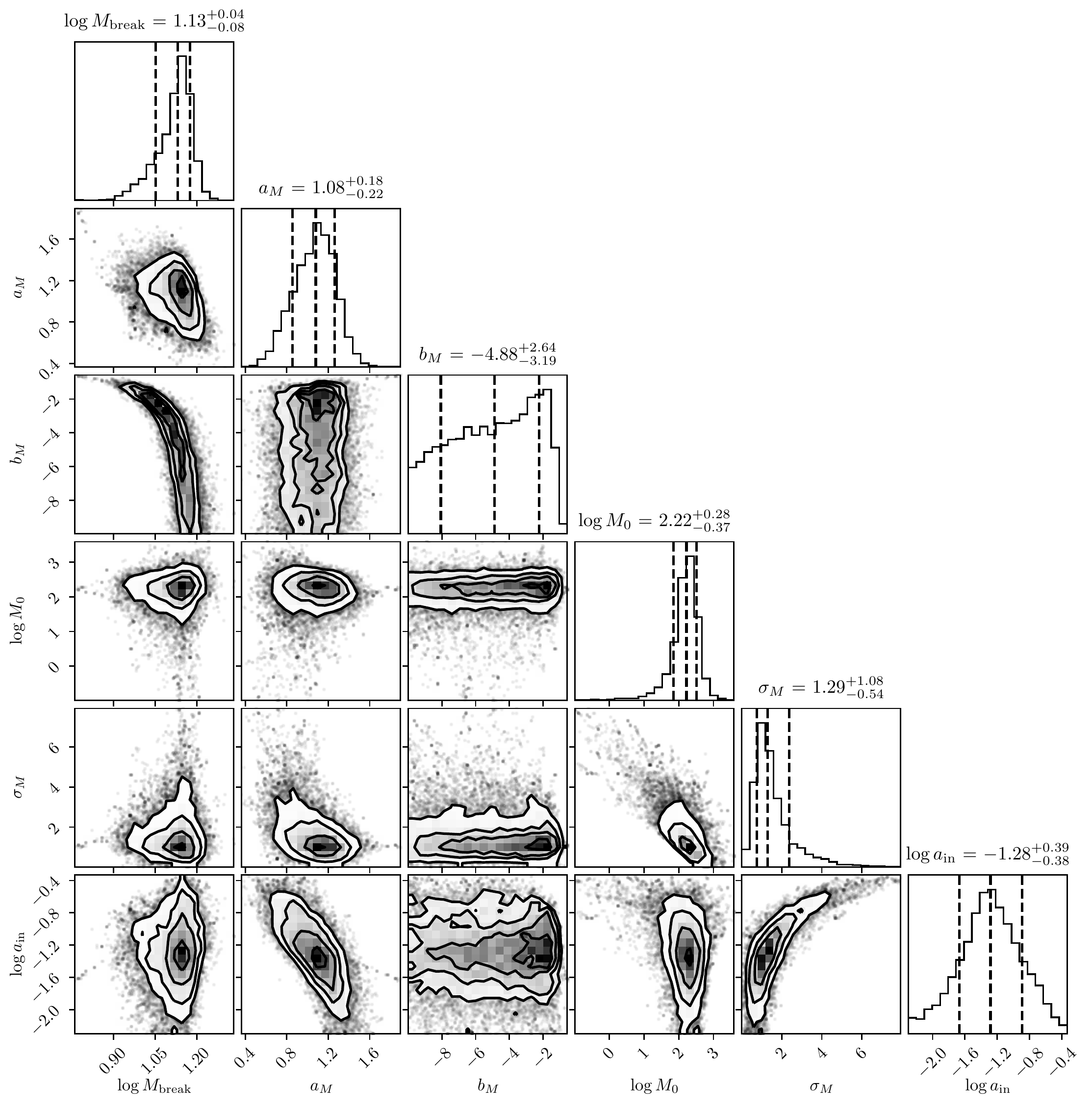}
    \caption{As in Figure~\ref{fig:fixpl_posterior} except for an alternative planet mass-radius relation (\textsc{Fix-PL-rg}).}
    \label{fig:fixplf17_posterior}
\end{figure}

As in Section~\ref{app:fixpl}, except we have adopted an alternative mass-radius relation with a different transition between rocky and gaseous as described in Section~\ref{sec:MR_relation} (equation~\ref{eq:MR_Fulton}). The posteriors are shown in Figure~\ref{fig:fixplf17_posterior}.

\subsection{Fixed inner-radius (\textsc{Fix-a\_in})}

\begin{figure}
    \centering
    \includegraphics[width=0.95\columnwidth]{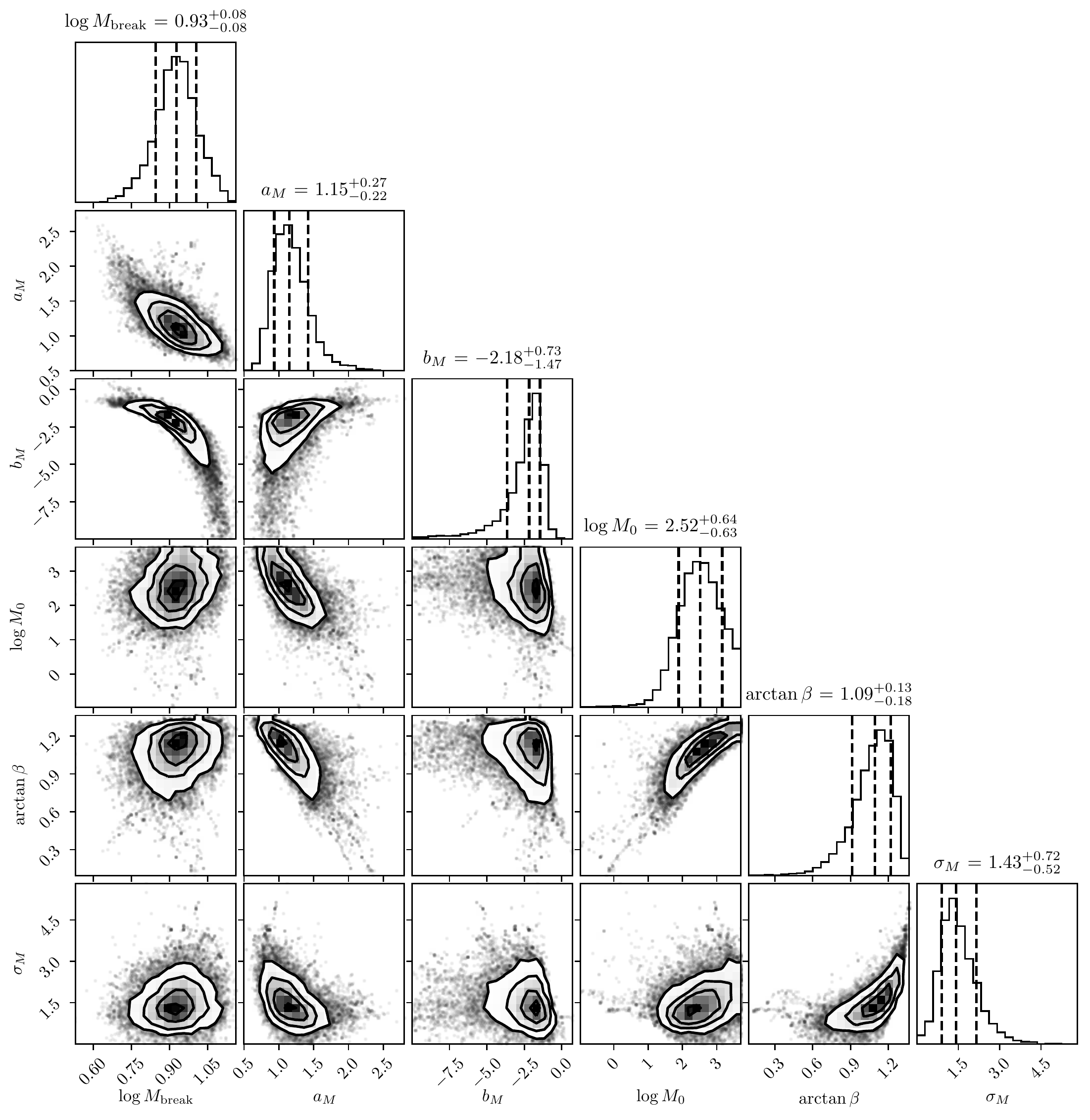}
    \caption{As in Figure~\ref{fig:fixpl_posterior} except for a fixed inner edge ($\log a_\mathrm{in}=-1.4$) and variable $\arctan \beta$ (\textsc{Fix-a\_in}).}
    \label{fig:fixain_posterior}
\end{figure}

The posterior as a result of fixing the inner-radius is shown in Figure~\ref{fig:fixain_posterior}, with the same priors as described above (Section~\ref{app:fixpl}), plus a uniform prior for $\theta$ between zero and $\pi$, where $\theta\equiv \arctan \beta$ for power-law $\beta$. In this case large $\beta$ (steep power-law) is associated with large $\sigma_M$. We also find that the steepest power-laws ($\theta\rightarrow \pi/2$) are effectively forbidden by the upper limit we place on $M_0$. To yield better constraints on $\beta$ we require additional data or a more restrictive and physically motivated prior.

\subsection{Double broken power-law model (\textsc{BPL-PR})}

\begin{figure}
    \centering
    \includegraphics[width=0.95\columnwidth]{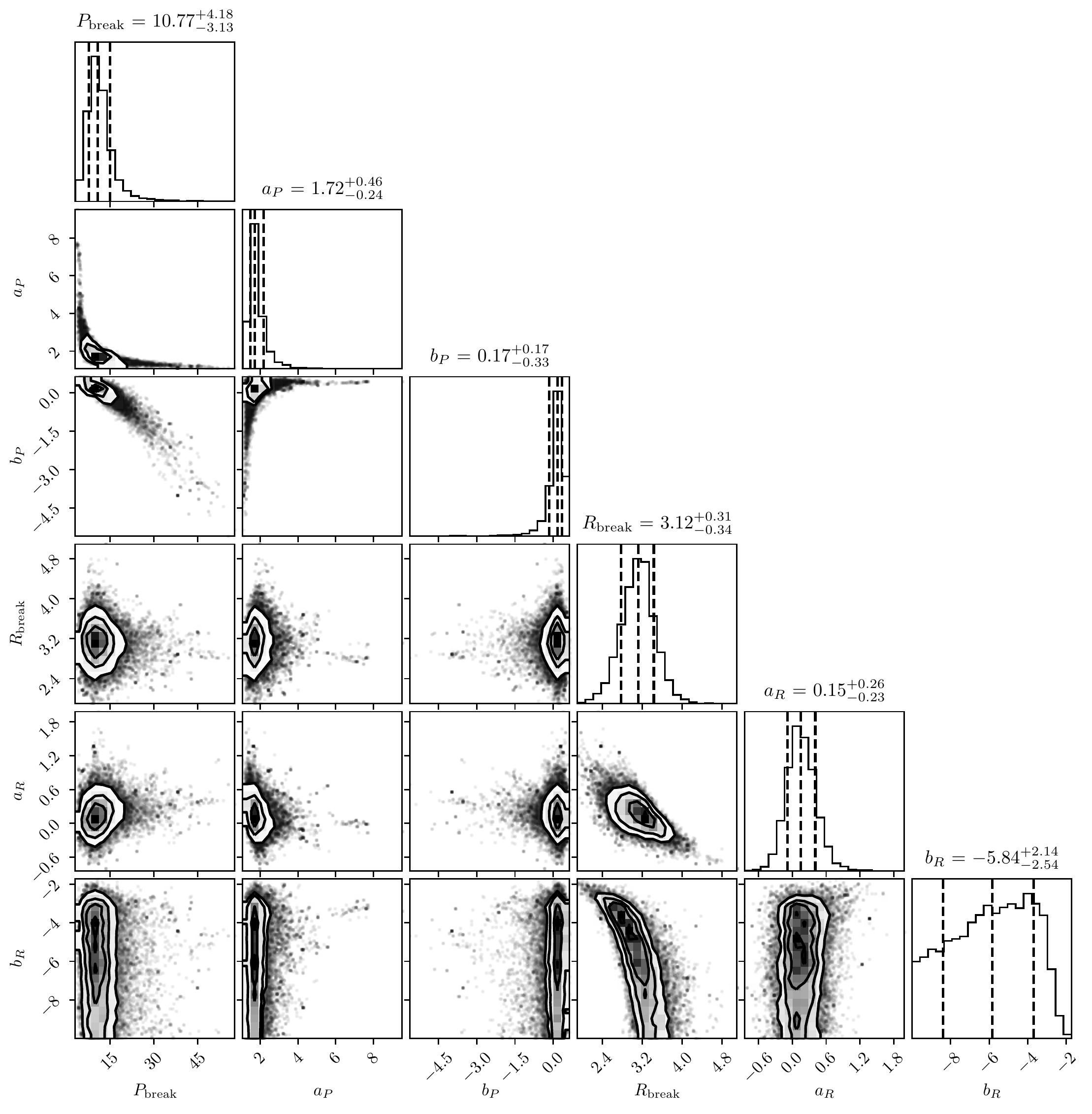}
    \caption{Posterior distributions resulting from the MCMC parameter space exploration for the double power-law PDF model in period and radius \citepalias[cf.][]{Mulders18}. The period break $P_\mathrm{break}$ is quoted in units of days and the radius break $R_\mathrm{break}$ in Earth radii, $R_\oplus$.  }
    \label{fig:simple_posterior}
    \label{lastpage}
\end{figure}

The posterior distribution for the fitting of the separable model in period and radius is shown in Figure~\ref{fig:simple_posterior}. The priors are uniform in the range $P_\mathrm{break}=1{-}100$~days, $R_\mathrm{break}=1{-}10\,R_\oplus$ and power-law indices $-10$ to $10$. The resulting distributions are similar to those inferred by \citetalias[][]{Mulders18}. 


\bsp	


 \end{document}